\def\Box{\kern1pt\vbox{\hrule height 1.2pt\hbox{\vrule width 1.2pt\hskip 3pt
   \vbox{\vskip 6pt}\hskip 3pt\vrule width 0.6pt}\hrule height 0.6pt}\kern1pt}
\def\gtwid{\mathrel{\raise.3ex\hbox{$>$\kern-.75em\lower1ex\hbox{$\sim$}}}}
\def\ltwid{\mathrel{\raise.3ex\hbox{$<$\kern-.75em\lower1ex\hbox{$\sim$}}}}
\def\be{\begin{equation}}
\def\ee{\end{equation}}
 
\documentstyle[epsfig,12pt]{article}

\begin{document}
\begin{titlepage}
\begin{flushright}
gr-qc/0205130 \\ UFIFT-HEP-02-15 \\ HD-THEP-02-13 \\ Imperial/TP/1-02/22
\end{flushright}
\vspace{.4cm}
\begin{center}
\textbf{One Loop Vacuum Polarization in a Locally de Sitter Background}
\end{center}
\begin{center}    
T. Prokopec$^*$
\end{center}
\begin{center}
\textit{Institute for Theoretical Physics, Heidelberg University, \\
Philosophenweg 16, D-69120 Heidelberg, Germany}
\end{center}
\begin{center}
O. T\"{o}rnkvist$^{\circ}$
\end{center}
\begin{center}
\textit{ Theoretical Physics Group, Imperial College \\
Prince Consort Road, London SW7 2BZ, U.K.} 
\end{center}
\begin{center}
R. P. Woodard$^{\ddagger}$
\end{center}
\begin{center}
\textit{Department of Physics, University of Florida \\ 
Gainesville, FL 32611 USA}
\end{center}
\begin{center}
ABSTRACT
\end{center}
\hspace*{.5cm} We compute the one loop vacuum polarization from massless, 
minimally coupled scalar QED in a locally de Sitter background. Gauge
invariance is maintained through the use of dimensional regularization,
whereas conformal invariance is explicitly broken by the scalar kinetic 
term as well as through the conformal anomaly. A fully renormalized result 
is obtained. The one loop corrections to the linearized, effective field 
equations do not vanish when evaluated on-shell. In fact the on-shell one 
loop correction depends quadratically on the inflationary scale factor, 
similar to a photon mass.  The contribution from the conformal anomaly is 
insignificant by comparison.
\begin{flushleft}
\end{flushleft}
\vspace{.4cm}
\begin{flushleft}
$^*$ e-mail: T.Prokopec@thphys.uni-heidelberg.de \\
$^{\circ}$ e-mail: o.tornkvist@ic.ac.uk \\
$^{\ddagger}$ e-mail: woodard@phys.ufl.edu
\end{flushleft}
\end{titlepage}

\section{Introduction}

One of many exciting and potentially testable consequences of inflationary
cosmology is a mechanism for generating the primordial cosmic magnetic 
fields which may have served as the seeds for the currently observed galactic
field of about $10^{-6}~{\rm Gauss}$. The idea \cite{DDPT1,DDPT2} is that 
the inflationary production of light, minimally coupled, charged scalars 
--- such as the Higgs --- resulted in the photon acquiring a plasma mass 
of about $m_{\gamma} \sim e H$, where $H \sim 10^{12}~{\rm GeV}$ is the 
inflationary Hubble parameter. Of course this would suppress the creation 
of photons during inflation, but it would vastly amplify the zero point 
energy of the super-horizon modes,
\be
\frac12 \hbar \omega \longrightarrow \frac12 \sqrt{k^2 e^{-2Ht} 
+ m^2_{\gamma}} \; ,
\ee
where $k = 2\pi/\lambda$ is the co-moving wave number. After the end of 
inflation the charged plasma dissipates --- either by annihilation or
through being redshifted into insignificance. If this happens quickly 
enough the enormous zero point energies are shed as coherent ensembles
of very long wave length photons which would manifest as magnetic fields 
on super-horizon scales.

To be more quantitative let us model the spacetime geometry during
inflation as locally de Sitter. We can express the invariant element 
conveniently either in co-moving or conformal coordinates,
\be
ds^2 = -dt^2 + e^{2 H t} d\vec{x} \cdot d\vec{x} = a^2 \Bigl[ -d\eta^2
+ d\vec{x} \cdot d\vec{x} \Bigr]\; .
\ee
The conformal factor and the transformation which relate the two coordinate
systems are,
\be
a(\eta) = -{1 \over H \eta} = e^{H t} \; .
\ee
Gravity is a non-dynamical background. The dynamical variables are the
vector potential $A_{\mu}(x)$ and a complex scalar $\phi(x)$. Their
Lagrangian is,
\be
{\cal L} = -\frac14 F_{\mu \nu} F_{\rho \sigma} g^{\mu \rho} 
g^{\nu \sigma} \sqrt{-g} - (\partial_{\mu} - i e A_{\mu}) \phi^* 
(\partial_{\nu} + i e A_{\nu}) \phi g^{\mu \nu} \sqrt{-g} \; .
\ee
One way of understanding the mass generation mechanism is by appealing to
the result of Vilenkin and Ford for the coincidence limit of a free scalar 
in Bunch-Davies vacuum \cite{AVLF},
\be
\Bigl\langle \Omega \Bigl\vert \phi^*(x) \phi(x) \Bigr\vert \Omega 
\Bigr\rangle_{\tiny {\rm free}} = \left({H \over 2 \pi}\right)^2 
\Bigl\{ {\rm UV} + H t \Bigr\} \; . \label{oldcoinc}
\ee
Here ``UV'' stands for an ultraviolet divergent constant. If we infer an 
approximate action for the photons by replacing the various $\phi^* \phi$ 
terms by the finite part of their vacuum expectation values, the result 
seems to be a time-dependent photon mass,
\be
m^2_{\gamma} = {e^2 H^2 \over 2 \pi^2} H t = {e^2 H^2 \over 2 \pi^2} 
\ln(a) \; .
  \label{Hartree-Fock}
\ee

Although very suggestive, the analysis of the preceding paragraph is not 
really consistent quantum field theory. The kinematic properties of a 
particle are encoded in its 2-point 1PI (one-particle-irreducible) function. 
The 2-photon 1PI function is known as ``the vacuum polarization'' and the 
one loop contributions to it are depicted in Figures 1-3. Making the 
replacement,
\be 
-e^2 A_{\mu} A_{\nu} \phi^* \phi g^{\mu\nu} \sqrt{-g} \longrightarrow 
-e^2 A_{\mu} A_{\nu} \left({H \over 2\pi}\right)^2 \Bigl\{ {\rm UV} +
H t \Bigr\} g^{\mu \nu} \sqrt{-g} \; .
\ee
corresponds to including only the diagram of Fig.~1. The other two graphs
are the same order ($e^2$) in perturbation theory and there seems to be no
good reason for ignoring them. The diagram of Fig.~2 is required to make the 
vacuum polarization gauge invariant. And the graph of Fig.~3 is needed to 
absorb the ultraviolet divergence. 

The purpose of this paper is to compute all three diagrams in a consistent 
regularization and to demonstrate that they induce corrections to the 
photon wave function very like those of a photon mass. In Section 2 we
review the familiar results from flat space. This serves as a useful 
introduction to using dimensional regularization in position space and
establishes a crucial correspondence limit for checking the accuracy of our
subsequent work. Section 3 gives the scalar propagator in $D$-dimensional de 
Sitter space. In Section 4 we first reduce the vacuum polarization to
manifestly transverse form, then we renormalize it. In Section 5 we take the
result on-shell to demonstrate that the tree order wave functions receive
one loop corrections like those of a photon mass. We discuss the result in
Section 6, giving special emphasis to the important issues which are still
open. A fuller discussion of what our result means physically can be found
in another work~\cite{letter}.

\begin{figure}[htbp]
\begin{center}
\epsfig{file=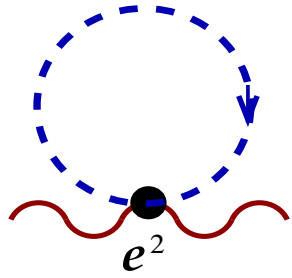, height=1.4in,width=1.5in}
\end{center}
\vskip -0.1in
\caption{\small One loop contribution to the vacuum polarization from the
4-point (seagull) interaction.}
\end{figure}

\begin{figure}[htbp]
\begin{center}
\epsfig{file=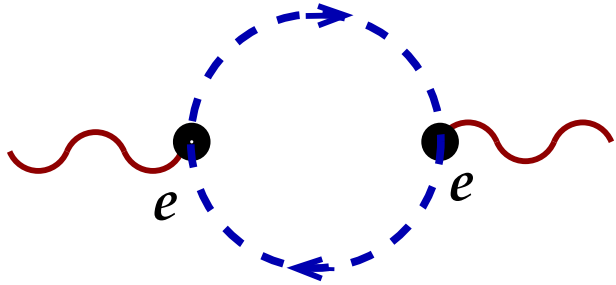, height=1.4in,width=3.2in}
\end{center}
\vskip -0.1in
\caption{\small One loop contribution to the vacuum polarization from two
3-point interactions.}
\end{figure}

\begin{figure}[htbp]
\begin{center}
\epsfig{file=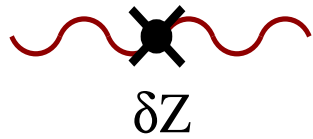, height=.7in,width=1.6in}
\end{center}
\vskip -0.1in
\caption{\small Photon field strength renormalization counterterm.}
\end{figure}

\section{Vacuum polarization in flat space}

\subsection{Momentum space treatment}

Although working in a curved background typically requires a position space 
treatment, it is desirable to begin our discussion of the vacuum polarization
with the traditional momentum space treatment. The contribution from the 
4-point interaction is depicted in Fig.~1. It actually vanishes for massless 
scalar QED in dimensional regularization,
\be
\Bigl({\rm Fig.~1}\Bigr)^{\mu \nu} = -2 i e^2 \eta^{\mu \nu} \int 
{d^Dk \over (2 \pi)^D} {-i \over k^2 - i \epsilon} = 0 \; .
\ee
Fig.~2 shows the contribution from two 3-point interactions,
\begin{eqnarray}
\Bigl({\rm Fig.~2}\Bigr)^{\mu \nu} & = & \int {d^Dk \over (2 \pi)^D} 
i e (p + 2 k)^{\mu} {-i \over (p + k)^2 - i \epsilon} i e (p + 2 k)^{\nu}
{-i \over k^2 - i \epsilon} \; , \\
& = & {i 2 e^2 \over (4 \pi)^{\frac{D}2} } {\Gamma(1 - \frac{D}2)
\Gamma^2(\frac{D}2) \over \Gamma(D)} (p^2)^{\frac{D}2 - 2} (p^2 
\eta^{\mu \nu} - p^{\mu} p^{\nu}) \; .
\end{eqnarray}
Since we have the result for any spacetime dimension $D$ we can note that its
vanishing on-shell (i.e., at $p^2 = 0$ in the transverse direction) for $D=4$ 
indicates that the photon stays massless in that dimension. The fact that the 
vacuum polarization does not vanish on-shell for $D=2$ indicates that the 
massless scalars form a massive photon bound state in two dimensions. This is 
a scalar version of the Schwinger model \cite{JS} in which the photon develops 
a mass of $m_{\gamma} \sim e$.

It is straightforward to extract the divergence as $D \equiv 4 - \epsilon$
approaches 4,
\be
\Bigl({\rm Fig.~2}\Bigr)^{\mu \nu} = - {i e^2 \over 24 \pi^2} \Bigl(p^2 
\eta^{\mu \nu} - p^{\mu} p^{\nu}\Bigr) {1 \over \epsilon} + {\rm finite} \; .
\ee
It is of course removed by the field strength renormalization of Fig.~3,
\be
\Bigl({\rm Fig.~3}\Bigr)^{\mu \nu} = - i {\delta Z} (\eta^{\mu \nu} p^2
- p^{\mu} p^{\nu}) \; .
\ee
One sets $\Delta Z = - e^2 \mu^{-\epsilon}/({24 \pi^2 \epsilon})$ plus any
convenient finite term.

\subsection{Position space treatment}

Let us now repeat the exercise in position space. The propagator is,
\be
i \Delta(x;x') = {\Gamma(1 - \frac{\epsilon}2) \over 4 \pi^{2 - 
\frac{\epsilon}2} } {1 \over ({\Delta x}^2)^{1 - \frac{\epsilon}2} } \; ,
\ee
where we define $\Delta x^2$ to be,
\be
{\Delta x}^2 \equiv \Vert \vec{x} - \vec{x}' \Vert^2 - (\vert t - t'\vert
- i \delta)^2 \; .
\ee
We will use the same figures for the position space diagrams.  Of course the 
first diagram still gives zero,
\be
\Bigl({\rm Fig.~1}\Bigr)^{\mu \nu} = -2 i e^2 \eta^{\mu \nu} i \Delta(x;x)
\delta^D(x - x') = 0 \; . \label{zero}
\ee
Note the general rule for using dimensional regularization in position space:
{\it the parameter $\epsilon$ is assumed to lie in the range for which terms
such as} ${\Delta x}^{ -N + \epsilon}$ {\it vanish at coincidence}. For
example, to derive (\ref{zero}) we assume $\epsilon > 2$, even though we shall
eventually take $\epsilon$ to zero. 

The position space versions of the other two diagrams are,
\be
\Bigl({\rm Fig.~2}\Bigr)^{\mu \nu} = 2 e^2 \eta^{\mu \rho} \eta^{\nu \sigma}
\Bigl[ \partial_{\rho} i \Delta(x;x') \partial_{\sigma}' i \Delta(x;x') -
i \Delta(x;x') \partial_{\rho} \partial_{\sigma}' i \Delta(x;x') \Bigr]
\; . \label{Fig2pos}
\ee
\be
\Bigl({\rm Fig.~3}\Bigr)^{\mu \nu} = - i {\delta Z} (\eta^{\mu \nu} 
\partial' \cdot \partial - \partial^{\prime \mu} \partial^{\nu}) 
\delta^D(x - x') \; .
\ee
Note that these particular diagrams involve no integrations when written in
position space. At two loop order there are integrations.

The first step in evaluating (\ref{Fig2pos}) is to substitute the propagator
and take the derivatives,
\begin{eqnarray}
i \Bigl[\mbox{}^{\mu}\Pi^{\nu}_{1+2}\Bigr](x;x') & = & 2 e^2 \left({\Gamma(1 -
\frac{\epsilon}2) \over 4 \pi^{2 - \frac{\epsilon}2}}\right)^2 \left[
-(2 - \epsilon)^2 {{\Delta x}^{\mu} {\Delta x}^{\nu} \over {\Delta x}^{8
- 2 \epsilon}} \right. \nonumber \\
& & \hspace{.5cm} \left.  - (2 - \epsilon) {\eta^{\mu \nu} \over 
{\Delta x}^{6 - 2 \epsilon}} + (2 - \epsilon) (4 - \epsilon) 
{{\Delta x}^{\mu} {\Delta x}^{\nu} \over {\Delta x}^{8 - 2 \epsilon}} 
\right] \; , \\
& = & {e^2 \over 8 \pi^4} \pi^{\epsilon} \Gamma^2(1 - {\textstyle
\frac{ \epsilon}2}) (2 - \epsilon) \left[{- \eta^{\mu \nu} \over 
{\Delta x}^{6 - 2 \epsilon}} + {2 {\Delta x}^{\mu} {\Delta x}^{\nu} \over 
{\Delta x}^{8 - 2 \epsilon}} \right] \; . \label{flatPi}
\end{eqnarray}
We anticipate the notation of curved space in denoting the vacuum polarization
as a bi-tensor function, $\Bigl[\mbox{}^{\mu}\Pi^{\nu}_{1+2}\Bigr](x;x')$.
This means that the first index ($\mu$) transforms according to the tangent 
space at the first argument ($x$) and the same relative relation exists
between the second index ($\nu$) transforms and the second argument ($x'$).

The next step is reaching manifestly transverse form. This is done by writing
inverse powers of ${\Delta x}$ as derivatives of lower powers. For example, 
one can easily derive the identity,
\be 
\partial^{\prime \mu} \partial^{\nu} {1 \over {\Delta x}^{4 - 2 \epsilon}}
= - (4 - 2 \epsilon) \left[{- \eta^{\mu \nu} \over {\Delta x}^{6 - 2 
\epsilon}} + (6 - 2 \epsilon) {{\Delta x}^{\mu} {\Delta x}^{\nu} \over
{\Delta x}^{8 - 2 \epsilon}} \right] \; .
\ee
By combining with $\eta^{\mu\nu}$ times the trace it follows that,
\be
\left[{ -\eta^{\mu \nu} \over {\Delta x}^{6 - 2 \epsilon}} + {2 
{\Delta x}^{\mu} {\Delta x}^{\nu} \over {\Delta x}^{8 - 2 \epsilon}}
\right] = {1 \over 2 (2 - \epsilon) (3 - \epsilon)} \Bigl[ \eta^{\mu \nu}
\partial' \cdot \partial - \partial^{\prime \mu} \partial^{\nu} \Bigr]
{1 \over {\Delta x}^{4 - 2 \epsilon}} \; .
\ee
Substitution into (\ref{flatPi}) gives a manifestly transverse form,
\be
i \Bigl[\mbox{}^{\mu}\Pi^{\nu}_{1+2}\Bigr](x;x') = {e^2 \over 16 \pi^4} 
{\pi^{\epsilon} \Gamma^2(1 - \frac{\epsilon}2) \over 3 - \epsilon} \Bigl[
\eta^{\mu \nu} \partial' \cdot \partial - \partial^{\prime \mu} 
\partial^{\nu} \Bigr] {1 \over {\Delta x}^{4 - 2 \epsilon}} \; .
\ee

The next step is extracting the ultraviolet divergence. This typically comes
from a term of the form $1/{\Delta x}^{4 - 2 \epsilon}$ through the identity,
\be
\partial^2 {1 \over {\Delta x}^{2 - 2 \epsilon}} = {-2 \epsilon (1 - 
\epsilon) \over {\Delta x}^{4 - 2 \epsilon}} \; . \label{firstd}
\ee
The reader will note that the various derivatives have so far failed to
induce any delta functions. That was because there were either too few 
derivatives or because the power of ${\Delta x}$ was wrong. Getting a delta
function in dimensional regularization  requires two derivatives acting on 
precisely the power ${\Delta x}^{\epsilon - 2}$,
\be
\partial^2 {1 \over {\Delta x}^{2 - \epsilon}} = {4 i \pi^{2 - \frac{
\epsilon}2} \over \Gamma(1 - \frac{\epsilon}2) } \delta^D(x - x') \; .
\label{secd}
\ee
Combining (\ref{firstd}) and (\ref{secd}) allows us to extract the ultraviolet
divergence in a form which can be canceled by a local counterterm,
\begin{eqnarray}
\lefteqn{{1 \over {\Delta x}^{4 - 2 \epsilon}} = {-\partial^2 \over 2 \epsilon
(1 - \epsilon)} {1 \over {\Delta x}^{2 - 2 \epsilon}} \; , } \\
& & = - {\partial^2 \over 2 \epsilon (1 - \epsilon)} \left[{1 \over 
{\Delta x}^{2 - 2 \epsilon}} - {\mu^{-\epsilon} \over {\Delta x}^{2 - 
\epsilon}} \right] - {2 \pi^2 i (\sqrt{\pi} \mu)^{-\epsilon} \over \epsilon 
(1 - \epsilon) \Gamma(1 - \frac{\epsilon}2)} \delta^D(x - x') , \\
& & \longrightarrow - {\partial^2 \over 4} \left[ {\ln(\mu^2 {\Delta x}^2) 
\over {\Delta x}^2} \right] - {2 \pi^2 i (\sqrt{\pi} \mu)^{-\epsilon} \over 
\epsilon (1 - \epsilon) \Gamma(1 - \frac{\epsilon}2)} \delta^D(x - x') \; .
\label{UVid}
\end{eqnarray}
As usual, dimensional regularization has resulted in a scale $\mu$.

Substituting (\ref{UVid}) reveals the divergence structure we already found
in momentum space,
\begin{eqnarray}
\lefteqn{i \Bigl[\mbox{}^{\mu}\Pi^{\nu}_{1+2}\Bigr](x;x') \longrightarrow 
- {e^2 \over 192 \pi^4} \Bigl[ \eta^{\mu \nu} \partial' \cdot \partial 
- \partial^{\prime \mu} \partial^{\nu} \Bigr] \partial^2 \left[ {\ln(\mu^2 
{\Delta x}^2) \over {\Delta x}^2}\right] } \nonumber \\
& & -{i e^2 \over 8 \pi^2} \Bigl({\pi \over \mu^2}\Bigr)^{\frac{\epsilon}2} 
{\Gamma(1 - \frac{\epsilon}2 ) \over \epsilon (1 - \epsilon) (3 - \epsilon)} 
\Bigl[ \eta^{\mu \nu} \partial' \cdot \partial - \partial^{\prime \mu} 
\partial^{\nu} \Bigr] \delta^D(x - x') \; , \label{flattransPi} \\
& & \longrightarrow - {i e^2 \over 24 \pi^2} \Bigl[ \eta^{\mu \nu} 
\partial' \cdot \partial - \partial^{\prime \mu} \partial^{\nu} \Bigr] 
\delta^D(x - x') {1 \over \epsilon} + {\rm finite} \; . 
\end{eqnarray}
The cleanest subtraction in position space is to absorb the entire local term
by choosing $\delta Z$ to be,
\be
{\delta Z} = - {e^2 \over 8 \pi^2} \left({\pi \over \mu^2}\right)^{\frac{
\epsilon}2} {\Gamma(1 - \frac{\epsilon}2) \over \epsilon (1 - \epsilon) 
(3 - \epsilon)} \; .
\ee
With this choice the fully renormalized vacuum polarization becomes,
\be
i \Bigl[\mbox{}^{\mu}\Pi^{\nu}_{\tiny \rm ren}\Bigr](x;x') = - {e^2 \over 
192 \pi^4} \Bigl[ \eta^{\mu \nu} \partial' \cdot \partial - \partial^{\prime 
\mu} \partial^{\nu} \Bigr] \partial^2 \left[ {\ln(\mu^2 {\Delta x}^2) \over 
{\Delta x}^2}\right] \; .
\ee

\subsection{Going on-shell in position space}

The position space derivation we have just completed was simpler than its
momentum space cousin because no integrations had to be performed. They occur
when one goes on-shell. What we are really checking, when we take the vacuum
polarization on-shell, is whether or not there are quantum corrections to the
linearized wave functions. One might define these as the matrix element of
the operator $A_{\mu}(x)$ between the vacuum and a plane wave photon state.
The equations obeyed by this matrix element come from varying and linearizing
the effective action,
\begin{eqnarray}
\lefteqn{ \Gamma[A] = -\frac14 \int d^4x F_{\mu\nu} F^{\mu \nu} } \nonumber \\
& & + \frac12 \int d^4x \int d^4x' A_{\mu}(x) \Bigl[\mbox{}^{\mu}\Pi^{\nu}
\Bigr](x;x') A_{\nu}(x') + O(A^4) \; .
\end{eqnarray}
The associated field equations are,
\be
{\delta \Gamma[A] \over \delta A_{\mu}(x)} = \partial_{\nu} F^{\nu \mu}(x) 
+ \int d^4x' \Bigl[\mbox{}^{\mu}\Pi^{\nu} \Bigr](x;x') A_{\nu}(x') + O(A^3) 
= 0 \; .
\ee

The general classical solution comes from superposing plane waves of the form,
\be
A_{\mu}^0(x) = \epsilon_{\mu}(\vec{k}) e^{i k \cdot x} \; ,
\ee
where $k^0 = \Vert \vec{k} \Vert$ and the Lorentz gauge polarization
vectors obey $\epsilon_0 = 0 = k \cdot \epsilon$. To see if there are quantum
corrections one merely expands the solution in powers of $\hbar$,
\be
A_{\mu}(x) = A^0_{\mu}(x) + A^1_{\mu}(x) + \dots \; .
\ee
and then segregates all terms of the same order. Potential one loop 
corrections are determined by the equation,
\be
\Bigl[ \partial^2 \eta^{\mu \nu} - \partial^{\mu} \partial^{\nu} \Bigl]
A^1_{\nu}(x) = - \int d^4x' \Bigl[\mbox{}^{\mu}\Pi^{\nu} \Bigr](x;x') 
A^0_{\nu}(x') \; .
\ee
We therefore conclude that {\it the necessary and sufficient condition for 
there to be one loop corrections to the photon wave function is a nonvanishing
integral for the one loop vacuum polarization against a classical plane wave 
solution.}

\section{Scalar propagator in de Sitter $D$-space}

The behavior of free, massless and minimally coupled scalars on a locally de 
Sitter background has been investigated extensively \cite{AVLF,BAAF,LFLP,FSW}. 
Among the curious properties of these particles are the absence of 
normalizable, de Sitter invariant states \cite{BAAF} and the appearance of 
acausal infrared singularities when the Bunch-Davies vacuum is used with 
infinite spatial surfaces \cite{LFLP,FSW}. To regulate this infrared problem 
we work on the manifold $T^{D-1} \times R$, with the spatial coordinates in 
the finite range, $-H^{-1}/2 < x^i \leq H^{-1}/2$.  Although the actual 
propagator is a mode sum on this manifold, the small possible variation in 
conformal coordinates renders the first term of the Euler-Maclaurin formula 
--- just the integral --- an excellent approximation. So the finite spatial 
range of $T^{D-1}$ serves merely to cut off what would have been a logarithmic 
infrared divergence on $R^{D-1}$. In $D = 3 + 1$ spacetime dimensions the 
result is \cite{TsWo1},
\begin{equation}
{i \Delta}(x;x')\Bigl\vert_{D=4} = \left({H \over 2 \pi} \right)^2 \left\{ 
\frac1{y(x;x')} - \frac12 \ln\Bigl(y(x;x')\Bigr) + \frac12 \ln\Bigl(
a(\eta) a(\eta') \Bigr) \right\} \; , \label{D=4prop}
\end{equation}
where the modified de Sitter length function has the definition,\footnote{What 
is termed ``the de Sitter length function'' in the literature is,
\begin{eqnarray}
z(x;x') = 1 - y(x;x') \; . \nonumber
\end{eqnarray}
The geodesic length from $x^{\mu}$ to $x^{\prime \mu}$, $\ell(x;x')$, is
related to $y(x;x')$ as follows,
\begin{eqnarray}
y(x;x') = \sin^2\left(\frac12 H \ell(x;x') \right) \; . \nonumber
\end{eqnarray}}
\begin{equation}
y(x;x') \equiv a(\eta) a(\eta') H^2 \left[ \Vert \vec{x} - \vec{x}' \Vert^2 - 
(\vert \eta - \eta' \vert - i \delta)^2 \right] \; . \label{dSL}
\end{equation}

Neglecting the higher order Euler-Maclaurin terms does not prevent 
(\ref{D=4prop}) from solving the correct differential equation. The higher
terms also drop out of quite complicated, nonlinear relations such as the 
Ward identity for the one loop graviton self-energy \cite{TsWo2}. We shall
therefore regard the technique as valid and confine ourselves to finding the 
appropriate generalization of (\ref{D=4prop}) to $D$ spacetime dimensions. 

We seek a function of $y(x;x')$ and the two conformal factors which obeys,
\be
\eta^{\mu\nu} {\partial \over \partial x^{\mu}} a^{D-2}(\eta) {\partial 
\over \partial x^{\nu}} {i \Delta}(x;x') = i \delta^D(x-x') \; .
\ee
When the kinetic operator acts on a function of just $y(x;x')$ one finds,
\begin{eqnarray}
\lefteqn{\eta^{\mu\nu} {\partial \over \partial x^{\mu}} a^{D-2}(\eta) 
{\partial \over \partial x^{\nu}} f\Bigl(y(x;x')\Bigr) } \nonumber \\
& & = H^2 a^D(\eta) \Bigl\{ (4 y - y^2) f^{\prime\prime}(y) + D (2 - y) 
f'(y) - 4 i \delta(\eta - \eta') f'(y) \delta \Bigr\} . \qquad
\end{eqnarray}
The only symmetric function of $a(\eta)$ and $a(\eta')$ which can give the
same prefactor of $a^D(\eta)$ is a constant times the same logarithm that 
appears in (\ref{D=4prop}). The $D$-dimensional propagator must therefore 
take the form,
\be
{i\Delta(x;x')} = f\Bigl( y(x;x')\Bigr) + b \ln\Bigl( a(\eta) a(\eta')
\Bigr) \; ,
\ee
where the function $f(y)$ obeys,
\begin{eqnarray}
\lefteqn{H^2 a^D(\eta) \Bigl\{ (4 y - y^2) f^{\prime\prime}(y) + D (2 - y) 
f'(y) } \nonumber \\
& & - 4 i \delta(\eta - \eta') f'(y) \delta - b (D-1) \Bigr\} = i 
\delta^D(x - x') \; . \label{propeq}
\end{eqnarray}

The delta function (for $\delta \rightarrow 0$) obviously descends from a 
factor of $y^{1 - \frac{D}2}$. Series solution of the equation then generates 
an infinite sum of higher powers. Defining $D \equiv 4 - \epsilon$ and 
normalizing correctly gives,
\be
{H^{2 - \epsilon} \over 4 \pi^{2 - \frac{\epsilon}2} } 
\Gamma\left(1 - \frac{\epsilon}2\right) \left\{ {1 \over y^{1 - \frac{
\epsilon}2}} - \left(1 - \frac{\epsilon}2 \right) \sum_{n = 0}^{\infty}
{1 \over n + \frac{\epsilon}2} {\Gamma(3 + n - \frac{\epsilon}2) \over 
(n+1)! \Gamma(2 - \frac{\epsilon}2)} {y^{n + \frac{\epsilon}2} \over 4^{n+1}} 
\right\} \; .
\ee
This series solves (\ref{propeq}) for $b=0$, but it does not reduce to 
(\ref{D=4prop}) for $\epsilon = 0$. The $n=0$ term of the sum is not even 
finite in this limit! The resolution to both problems is a series of strictly 
nonnegative integer powers of $y$, which cancels the divergence and the 
unwanted terms. This series obeys the homogeneous equation up to a constant
which is canceled by the $b (D-1)$ term,
\begin{eqnarray}
\lefteqn{{i \Delta}(x;x') = \left({H \over 2 \pi}\right)^2 \left({H \over
\sqrt{\pi}} \right)^{-\epsilon} \Gamma\left(1 - \frac{\epsilon}2\right)
\left\{ {1 \over y^{1 - \frac{\epsilon}2}} + \left(1 - \frac{\epsilon}2 \right)
\left(1 - \frac{\epsilon}4 \right) \left({1 - y^{\frac{\epsilon}2} \over
\epsilon} \right) \right. } \nonumber \\
& & \hspace{1cm} + \left(1 - \frac{\epsilon}2 \right) \sum_{n = 1}^{\infty}
\left[\frac1{n} {\Gamma(3 + n - \epsilon) \over \Gamma(2 + n - \frac{\epsilon
}2)} - {1 \over n + \frac{\epsilon}2} {\Gamma(3 + n - \frac{\epsilon}2) \over 
(n+1)! \Gamma(2 - \frac{\epsilon}2)} y^{\frac{\epsilon}2} \right] {y^n \over
4^{n+1}} \nonumber \\
& & \hspace{6.5cm} \left. + \frac14 {\Gamma(3 - \epsilon) \over \Gamma(1 -
\frac{\epsilon}2)} \ln\Bigl( a(\eta) a(\eta') \Bigr) \right\} \; . \label{prop}
\end{eqnarray}

The great advantage of this regularization is that it preserves general
coordinate invariance (once $\delta$ is taken to zero). One might think that 
the propagator is unwieldy but this is not so in practice. For example, this
formalism has recently been used to compute and renormalize all {\it two loop}
contributions to the stress-energy tensor of a real scalar with a $\phi^4$ 
self-interaction \cite{OW}. The really cumbersome part of (\ref{prop}) is the 
infinite sum on the second line. But these terms all vanish at coincidence 
($y(x;x) = 0$) and they vanish for all $y(x;x')$ at $D=4$. So one need only 
retain them when they multiply something else that diverges like $1/\epsilon$. 
Note also that one need never worry about large $y(x;x')$ on account of 
causality.

All valid regularizations must reproduce the result of Vilenkin and Ford
that the coincidence limit of the propagator contains a finite term which
grows like $\ln(a) = H t$ \cite{AVLF}. To check this note that $y(x;x) 
=0$ at coincidence. When a variable vanishes like this in dimensional 
regularization one must always assume $\epsilon$ to be large enough that
the variable is raised to only nonnegative powers. We therefore find,
\begin{equation}
\lim_{x' \rightarrow x} {i\Delta}(x;x') = \left({H \over 2 \pi}\right)^2
\left({H \over \sqrt{\pi}}\right)^{-\epsilon} \left\{ \frac1{2 \epsilon}
\Gamma(3 - {\textstyle \frac{\epsilon}2} ) + \frac12 \Gamma(3 - \epsilon) 
\ln\Bigl(a(\eta)\Bigr) \right\} \; . \label{coinc}
\end{equation}
Note that (\ref{coinc}) is exact for arbitrary $\epsilon$ and indeed
reduces to (\ref{oldcoinc}) in the limit when $\epsilon$ approaches zero.

\section{Vacuum polarization in de Sitter}

The diagrams which contribute to the vacuum polarization in de Sitter 
background are drawn exactly as those of flat space, so we shall use the 
same figures. Of course there are now some factors of the de Sitter metric!
We express this in conformal coordinates and adopt the usual convention that 
indices are raised and lowered by the Lorentz metric,
\begin{eqnarray}
\Bigl({\rm Fig.~1}\Bigr)^{\mu \nu} & = & -2 i e^2 \sqrt{-g(x)} g^{\mu \nu}(x) 
i \Delta(x;x) \delta^D(x - x') \; , \\
& = & -2 i e^2 a^{D-2} \eta^{\mu\nu} i \Delta(x;x) \delta^D(x - x') \; .
\label{Fig1dS}
\end{eqnarray}
\begin{eqnarray}
\lefteqn{\Bigl({\rm Fig.~2}\Bigr)^{\mu \nu} = 2 e^2 \sqrt{-g(x)} g^{\mu 
\rho}(x) \sqrt{-g(x')} g^{\nu \sigma}(x') } \nonumber \\
& & \hspace{2cm} \times \Bigl[ \partial_{\rho} i \Delta(x;x') \partial_{
\sigma}' i \Delta(x;x') - i \Delta(x;x') \partial_{\rho} \partial_{\sigma}' i 
\Delta(x;x') \Bigr] \; , \label{Fig2dSa} \\
& & = 2 e^2 a^{D-2} a^{\prime D-2} \Bigl[ \partial^{\mu} i \Delta(x;x') 
\partial^{\prime \nu} i \Delta(x;x') - i \Delta(x;x') \partial^{\mu} 
\partial^{\prime \nu} i \Delta(x;x') \Bigr] \; . \qquad \label{Fig2dS}
\end{eqnarray}
\begin{eqnarray}
\Bigl({\rm Fig.~3}\Bigr)^{\mu \nu} & = & -i {\delta Z} \partial_{\rho} 
\Bigl(\sqrt{-g} \left[g^{\mu \nu} g^{\rho \sigma} - g^{\mu \sigma} g^{\nu \rho} 
\right] \partial^{\prime}_{\sigma} \delta^D(x - x') \Bigr)\; , \\
& = & - i {\delta Z} \left[ \eta^{\mu \nu} \partial' \cdot \partial - 
\partial^{\prime \mu} \partial^{\nu} \right] a^{D-4} \delta^D(x - x') \; .
\label{Fig3dS}
\end{eqnarray}
In these and subsequent expressions we define $a \equiv a(\eta)$ and $a'
\equiv a(\eta')$. Note as well that the scalar propagator is the de Sitter 
one (\ref{prop}).

It is convenient to subsume the complicated, $\epsilon$-dependent constants 
which appear in the propagator (\ref{prop}),
\be
i \Delta(x;x') \equiv \alpha \Bigl\{\gamma\Bigl( y(x;x')\Bigr) + \beta 
\ln\Bigl( a(\eta) a(\eta')\Bigr) \Bigr\}\; . \label{propshort}
\ee
Comparison with (\ref{prop}) reveals,
\be
\alpha \equiv \left(\frac{H}{2\pi} \right)^2 \left({\pi \over H^2}\right)^{
\frac{\epsilon}2} \Gamma(2 - {\textstyle \frac{\epsilon}{2}})\; , \hskip 0.7cm 
\beta \equiv \frac14 \frac{\Gamma(3-\epsilon)}{\Gamma(2 - \frac{\epsilon}2)} 
\; ,
\ee
and,
\begin{eqnarray}
\lefteqn{\gamma(y) \equiv \frac1{1 - {\textstyle {\epsilon \over 2}}} 
\frac1{y^{1 - {\epsilon \over 2}}} + (1 - {\textstyle {\epsilon \over 4}}) 
\left(\frac{1 - y^{\frac{ \epsilon}{2}}}{ \epsilon}\right) } \nonumber \\
& & + \sum_{n =1}^{\infty} \Bigg[\frac1{n} {\Gamma(3 + n - \epsilon) \over 
\Gamma(2 + n - \epsilon)} \frac{y^n}{4^{n+1}} - {1 \over {n + \frac{
\epsilon}2}} {{\Gamma(3 + n - \frac{\epsilon}{2})} \over {(n+1)! \Gamma(2 - 
\frac{\epsilon}2)}}} {{y^{n + \frac{\epsilon}2} \over {4^{n+1}}}\Bigg]\; .
\end{eqnarray}
In this notation derivatives of the propagator (\ref{propshort}) can be 
written as,
\begin{eqnarray}
\partial_{\rho} i\Delta(x;x') & = & \alpha \left\{\gamma'(y) {\partial y \over
\partial x^{\rho}} + \beta H a(\eta) \delta^0_{~\rho} \right\} \; , \\
\partial^{\prime}_{\sigma} i\Delta(x;x') & = & \alpha \left\{\gamma'(y) 
{\partial y \over \partial x^{\prime \sigma}} + \beta H a(\eta') \delta^0_{~
\sigma} \right\} \; , \\
\partial_{\rho} \partial^{\prime}_{\sigma} i\Delta(x;x') & = & \alpha \left\{
\gamma''(y) {\partial y \over \partial x^{\rho}} {\partial y \over \partial 
x^{\prime \sigma}} + \gamma'(y) {\partial^2 y \over \partial x^{\rho} \partial
x^{\prime \sigma}} \right\} \; .
\end{eqnarray}
We can therefore express the portion of (\ref{Fig2dSa}) within the brackets
as,
\begin{eqnarray}
\lefteqn{\frac1{\alpha^2} \Bigl[ \partial_{\rho} i\Delta(x;x') \partial^{
\prime}_{\sigma} i\Delta(x;x') - i\Delta(x;x') \partial_{\rho} \partial^{
\prime}_{\sigma} i\Delta(x;x') \Bigr] = } \nonumber \\
& & \left[ \gamma^{\prime 2} - \gamma'' \Bigl( \gamma + \beta \ln(a a') \Bigr) 
\right] {\partial y \over \partial x^{\rho}} {\partial y \over \partial 
x^{\prime \sigma}}- \gamma' \Bigl[ \gamma + \beta \ln(a a') \Bigr] {\partial^2
y \over \partial x^{\rho} \partial x^{\prime \sigma}} \nonumber \\
& & \hspace{3cm} + \beta \gamma' H \left[ a \delta^0_{~\rho} {\partial y \over 
\partial x^{\prime \sigma}} + a' {\partial y \over \partial x^{\rho}} 
\delta^0_{~\sigma} \right] + \beta^2 H^2 a a' \delta^0_{~\rho} 
\delta^0_{~\sigma} \; . \quad \label{dDdD}
\end{eqnarray}

It is straightforward to differentiate the de Sitter length function 
(\ref{dSL}),
\begin{eqnarray}
{\partial y \over \partial x^{\rho}} & = & a(\eta) H \left[ y \delta^0_{~ \rho}
+ 2 a(\eta') H {\Delta x}_{\rho} + 2 i H a(\eta') {\rm sgn}(\eta - \eta')
\delta^0_{~\rho} \delta \right] \; , \quad \\
{\partial y \over \partial x^{\prime \sigma}} & = & a(\eta') H \left[ y 
\delta^0_{~\sigma} - 2 a(\eta) H {\Delta x}_{\sigma} - 2 i H a(\eta) 
{\rm sgn}(\eta - \eta') \delta^0_{~\sigma} \delta \right] \; , \quad \\
{\partial^2 y \over \partial x^{\rho} \partial x^{\prime \sigma}} & = & 
a a' H^2 \left[y \delta^0_{~\rho} \delta^0_{~ \sigma} - 2 a(\eta) 
\delta^0_{~\rho} H {\Delta x}_{\sigma} + 2 a(\eta') H {\Delta x}_{\rho} 
\delta^0_{~\sigma} - 2 \eta_{\rho \sigma} \right. \nonumber \\
& & \hspace{.5cm} \left.  - 2 i a(\eta) a(\eta') H^2 \vert \eta - \eta' \vert 
\delta^0_{~\rho} \delta^0_{~\sigma} \delta - 4 i \delta(\eta - \eta') 
\delta^0_{~\rho} \delta^0_{~\sigma} \delta \right] \; . \quad \label{2prop}
\end{eqnarray}
Gauge invariance requires taking $\delta$ to zero. Only the final order 
$\delta$ term can contribute in this limit, and then only when it multiplies 
$\gamma'$,
\be
-4 i a a' H^2 \delta(\eta - \eta') \lim_{\delta \rightarrow 0} \gamma'\Bigl( 
y(x;x') \Bigr) \delta = {i \over \alpha a^{D-2}} \delta^D(x-x') \; .
\ee
The net result is to subtract off the purely temporal components of 
(\ref{Fig1dS}). We accordingly combine this term with $\Big({\rm Fig.~1}
\Bigr)^{\mu\nu}$ to form,
\be
i \Bigl[\mbox{}^{\mu}\Pi^{\nu}_1\Bigr](x;x') \equiv - i e^2 \alpha a^{2 -
\epsilon} \overline{\eta}^{\mu\nu} \left\{ \left({2 - \frac{\epsilon}2 \over
\epsilon}\right) + {\Gamma(3 - \epsilon) \over \Gamma(2 - \frac{\epsilon}2)}
\ln(a) \right\} \delta^D(x - x') \; . \label{Pi1comb}
\ee
A bar over a tensor indicates its zero components have been removed,
\be
\overline{\eta}^{\mu\nu} \equiv \eta^{\mu\nu} + \delta^{\mu}_{~0} 
\delta^{\nu}_{~0} \qquad , \qquad \overline{\partial}^{\mu} \equiv \partial^{
\mu} - \delta^{\mu}_{~ 0} \partial^0 \; .
\ee
The left-over portion of $\Bigl({\rm Fig.~2}\Bigr)^{\mu\nu}$ will be known as
$i \Bigl[\mbox{}^{\mu}\Pi^{\nu}_2\Bigr](x;x')$.

Expression (\ref{dDdD}) seems complicated due to the infinite sum in
the definition of $\gamma(y)$. However, we need only retain terms which 
survive as $\epsilon\rightarrow 0$,
\begin{eqnarray}
\lefteqn{\gamma^{\prime 2} - \gamma'' \Bigl[\gamma + \beta \ln(a a') \Bigr] 
\longrightarrow - {1 \over 1 - \frac{\epsilon}2 } {1 \over y^{4 - \epsilon}} 
+ \left({2 - \frac{\epsilon}2 \over \epsilon} \right) {1 \over y^{3 - 
\epsilon}} } \nonumber \\
& & - {(2 - \frac{\epsilon}2)^2 \over 2 \epsilon} {1 \over y^{3- \frac{
\epsilon}2}} - \left({2 - \frac{\epsilon}2 \over 4}\right) {\Gamma(3 - 
\epsilon) \over \Gamma(2 - \frac{\epsilon}2)} {\ln(a a') \over y^{3 - \frac{
\epsilon}2}} + {\ln(H^2 {\Delta x}^2) \over 4 y^2} + {1 \over 4 y^2} 
\; , \qquad \\
\lefteqn{\gamma' \Bigl[\gamma + \beta \ln(a a') \Bigr] \longrightarrow - 
{1 \over 1 - \frac{\epsilon}2} {1 \over y^{3 - \epsilon}} + {(1 - \epsilon)
( 2 - \frac{\epsilon}2 ) \over 2 \epsilon (1 - \frac{\epsilon}2) } {1 
\over y^{2 - \epsilon}} } \nonumber \\
& & \hspace{2cm} - \left( {2 - \frac{\epsilon}2 \over 2 \epsilon} \right) 
{1 \over y^{2 - \frac{\epsilon}2 }} - \frac14 {\Gamma(3 - \epsilon) \over 
\Gamma(2 - \frac{ \epsilon}2)} {\ln(a a') \over y^{2 - \frac{\epsilon}2}} 
+ {\ln(H^2 {\Delta x }^2) \over 4 y} \; .
\end{eqnarray}
After taking $\epsilon$ to zero in other non-divergent terms the result is,
\begin{eqnarray}
\lefteqn{i \Bigl[\mbox{}^{\mu}\Pi^{\nu}_2\Bigr](x;x') = 2 e^2 \alpha^2 
a^{2-\epsilon} a^{\prime 2 - \epsilon} \left\{ 4 a^2 a^{\prime 2} H^4 
{\Delta x}^{\mu} {\Delta x}^{\nu} \left[ {1 \over 1 - \frac{\epsilon}2 } 
{1 \over y^{4 - \epsilon}} - \left( {2 - \frac{\epsilon}2 \over \epsilon
\, y^{3 - \epsilon}} \right) \right. \right. } \nonumber \\
& & \left. + {(2 - \frac{\epsilon}2)^2 \over 2 \epsilon} {1 \over y^{3- \frac{
\epsilon}2}} + \left({2 - \frac{\epsilon}2 \over 4}\right) {\Gamma(3 - 
\epsilon) \over \Gamma(2 - \frac{\epsilon}2)} {\ln(a a') \over y^{3 - \frac{
\epsilon}2}} - {\ln(H^2 {\Delta x}^2) \over 4 y^2} - {1 \over 4 y^2} \right]
\nonumber \\
& & - 2 a a' H^2 \eta^{\mu \nu} \left[ {1 \over 1 - \frac{\epsilon}2 } {1 
\over y^{3 - \epsilon}} - {(1 - \epsilon) (2 - \frac{\epsilon}2) \over 2
\epsilon (1- \frac{\epsilon}2) } {1 \over y^{2 - \epsilon}} + \left({2 - 
\frac{\epsilon}2 \over 2 \epsilon}\right) {1 \over y^{2 - \frac{\epsilon}2}} 
\right. \nonumber \\
& & \hspace{6.5cm} \left. + \frac14 {\Gamma(3 - \epsilon) \over \Gamma(2 - 
\frac{\epsilon}2)} {\ln(a a') \over y^{2 - \frac{\epsilon}2}} - {\ln(H^2 
{\Delta x}^2) \over 4 y} \right] \nonumber \\
& & + a a' H^2 \Bigl[- a' H {\Delta x}^{\mu} \delta^{\nu}_{~ 0} + a 
\delta^{\mu}_{~ 0} H {\Delta x}^{\nu}\Bigr] \left[{ \ln(H^2 {\Delta x}^2) + 
1 \over y^2} \right] \nonumber \\
& & \hspace{7cm} \left. + a a' H^2 \delta^{\mu}_{~ 0} \delta^{\nu}_{~ 0} 
{\ln(H^2 {\Delta x}^2) \over 2 y} \right\} \; , \\
\lefteqn{= -{ 4 e^2 \alpha^2 H^{2\epsilon - 4} \over 1 - \frac{\epsilon}2}
\left[ \eta^{\mu\nu} - 2 {{\Delta x}^{\mu} {\Delta x}^{\nu} \over {\Delta x}^2}
\right] {1 \over {\Delta x}^{6 - 2\epsilon}} } \nonumber \\
& & + 2 e^2 \alpha^2 H^{2 \epsilon - 2} {1 \over \epsilon} \left({2 - 
\frac{\epsilon}2 \over 1 - \frac{\epsilon}2} \right) a a' \left[(1 - \epsilon) 
\eta^{\mu\nu} - (4 - 2 \epsilon) {{\Delta x}^{\mu} {\Delta x}^{\nu} \over
{\Delta x}^2}\right] {1 \over {\Delta x}^{4 - 2 \epsilon}} \nonumber \\
& & - 2 e^2 \alpha^2 H^{\epsilon - 2} a^{1 - \frac{\epsilon}2} a^{\prime 1 -
\frac{\epsilon}2} \left[ \left({2 - \frac{\epsilon}2 \over \epsilon}\right)
+ \frac12 {\Gamma(3 - \epsilon) \over \Gamma(2 - \frac{\epsilon}2)} \ln(a a')
\right] \nonumber \\
& & \hspace{6cm} \times \left[ \eta^{\mu \nu} - (4 - \epsilon) 
{{\Delta x}^{\mu} {\Delta x}^{\nu} \over {\Delta x}^2} \right] {1 \over 
{\Delta x}^{4 - \epsilon}} \nonumber \\
& & + {e^2 H^2 \over 8 \pi^4} a^2 a^{\prime 2} \Bigl[- a^{-1} H {\Delta x}^{
\mu} \delta^{\nu }_{~ 0} + a^{\prime -1} \delta^{\mu}_{~ 0} H {\Delta x}^{\nu} 
- H^2 {\Delta x}^{\mu} {\Delta x}^{\nu} \Bigr] \nonumber \\
& & \hspace{1.5cm} \times \left[{ \ln(H^2 {\Delta x}^2) + 1 \over {\Delta x}^4}
\right] + {e^2 H^4 \over 16 \pi^4} a^2 a^{\prime 2} \Bigl[\eta^{\mu \nu} +
\delta^{\mu}_{~ 0} \delta^{\nu}_{~ 0} \Bigr] {\ln(H^2 {\Delta x}^2) \over 
{\Delta x}^2} \; . \label{Fig2B}
\end{eqnarray}

The next step is reducing to manifestly transverse form. The first term in
(\ref{Fig2B}) is exactly the same as the flat space result (\ref{flatPi}) and
its reduction gives (\ref{flattransPi}), as before. The second term is
reduced with the identity,
\begin{eqnarray}
\lefteqn{\left[ (1 - \epsilon) \eta^{\mu\nu} - (4 - 2 \epsilon) 
{{\Delta x}^{\mu} {\Delta x}^{\nu} \over {\Delta x}^2} \right] {1 \over 
{\Delta x}^{4 - 2 \epsilon}} =} \nonumber \\
& & \hspace{4cm} - {1 \over 2 - 2 \epsilon} \left[ \eta^{\mu \nu} \partial' 
\cdot \partial - \partial^{\prime \mu} \partial^{\nu} \right] {1 \over 
{\Delta x}^{2 - 2 \epsilon}} \; .
\end{eqnarray}
The fact that the third term goes like ${\Delta x}^{\epsilon - 4}$ means that
its reduction produces a local term,
\begin{eqnarray}
\lefteqn{ \left[ \eta^{\mu\nu} - (4 - \epsilon) {{\Delta x}^{\mu} {\Delta x}^{
\nu} \over {\Delta x}^2} \right] {1 \over {\Delta x}^{4 - \epsilon}} = }
\nonumber \\
& & - {1 \over 2 - \epsilon} \Bigl[ \eta^{\mu \nu} \partial' \cdot \partial 
- \partial^{\prime \mu} \partial^{\nu} \Bigr] {1 \over {\Delta x}^{ 2 - 
\epsilon}} -{2 i \pi^{2 - \frac{\epsilon}2} \over \Gamma(2 - \frac{\epsilon}2)}
\overline{\eta}^{\mu \nu} \delta^D(x - x') \; .
\end{eqnarray}
This local term completely cancels (\ref{Pi1comb}). The sum of all terms from
Figures 1 and 2 is therefore,
\begin{eqnarray}
\lefteqn{i \Bigl[\mbox{}^{\mu}\Pi^{\nu}_{1+2}\Bigr](x;x') = {e^2 \over 8 \pi^4}
\left\{ { \pi^{\epsilon} \Gamma^2(1 - \frac{\epsilon}2) \over 2 (3 - \epsilon)}
\Bigl[ \eta^{\mu \nu} \partial' \cdot \partial - \partial^{\prime \mu} 
\partial^{\nu} \Bigr] {1 \over {\Delta x}^{4 - 2 \epsilon}} \right.} 
\nonumber \\
& & \hspace{1cm} - {1 \over \eta \eta'} \Bigl[ \eta^{\mu \nu} \partial' \cdot 
\partial - \partial^{\prime \mu} \partial^{\nu} \Bigr] \left[{ \frac12 \ln(H^2 
{\Delta x}^2) + 1 \over {\Delta x}^2}\right] + \overline{\eta}^{\mu \nu} 
{\ln(H^2 {\Delta x}^2) \over 2 \eta^2 \eta^{\prime 2} {\Delta x}^2} 
\nonumber \\
& & \hspace{1.5cm} + \left. \Bigl[\eta {\Delta x}^{\mu} \delta^{\nu }_{~ 0} - 
\eta' \delta^{\mu}_{~ 0} {\Delta x}^{\nu} - {\Delta x}^{\mu} {\Delta x}^{\nu} 
\Bigr] \left[{\ln(H^2 {\Delta x}^2) + 1 \over \eta^2 \eta^{\prime 2} 
{\Delta x}^4} \right] \right\} \; . \qquad \label{Fig1+2}
\end{eqnarray}
Note that only the first term --- the one that survives in the flat space 
limit --- still requires regularization.

Although expression (\ref{Fig1+2}) is transverse this is not yet quite manifest
owing to the factor of $1/(\eta \eta')$ still standing to the left of the 
derivatives in the second term. After some tedious tensor algebra one
finds,
\begin{eqnarray}
\lefteqn{i \Bigl[\mbox{}^{\mu}\Pi^{\nu}_{1+2}\Bigr](x;x') = {e^2 \over 8 \pi^4}
\left\{ { \pi^{\epsilon} \Gamma^2(1 - \frac{\epsilon}2) \over 2 (3 - \epsilon)}
\Bigl[ \eta^{\mu \nu} \partial' \cdot \partial - \partial^{\prime \mu} 
\partial^{\nu} \Bigr] {1 \over {\Delta x}^{4 - 2 \epsilon}} \right.} 
\nonumber \\
& & \hspace{4cm} - \Bigl[ \eta^{\mu \nu} \partial' \cdot \partial - \partial^{
\prime \mu} \partial^{\nu} \Bigr] \left[{ \frac12 \ln(H^2 {\Delta x}^2) + 1 
\over \eta \eta' {\Delta x}^2}\right] \nonumber \\
& & \hspace{2cm} \left. + \Bigl[ {\overline{\eta}}^{\mu\nu} \vec{\nabla}' 
\cdot \vec{\nabla} - \overline{\partial}^{\prime \mu} \overline{\partial}^{\nu}
\Bigr] \left[{ \frac18 \ln^2(H^2 {\Delta x}^2) + \frac12 \ln(H^2 {\Delta x}^2) 
\over \eta^2 \eta^{\prime 2}} \right] \right\} \; . \qquad
\end{eqnarray}
It remains only to extract the divergence and subtract it with Fig.~3. 
Although the divergence is exactly the same as in flat space 
(\ref{flattransPi}) the counterterm is not, owing to the factor of $a^{D-4}$ 
in (\ref{Fig3dS}). The incomplete cancellation gives rise to a finite factor 
of $\ln(a)$ times a delta function. Of course this is the conformal anomaly 
\cite{Dolgov,SQED}. The renormalized vacuum polarization is,
\begin{eqnarray}
\lefteqn{i \Bigl[\mbox{}^{\mu}\Pi^{\nu}_{\rm ren}\Bigr](x;x') = {e^2 \over 8 
\pi^4} \left\{ - \Bigl[ \eta^{\mu \nu} \partial' \cdot \partial - \partial^{
\prime \mu} \partial^{\nu} \Bigr] \left[ \partial^2 \left( {\ln(\mu^2 {\Delta
x}^2) \over 24 {\Delta x}^2} \right) \right. \right. } \nonumber \\
& & \hspace{4cm} + \left. \left({\frac12 \ln(H^2 {\Delta x}^2) + 1 \over \eta 
\eta' {\Delta x}^2} \right) + {i \pi^2 \over 3} \ln(a) \delta^4(x - x') \right]
\nonumber \\
& & \hspace{2cm} \left. + \Bigl[ {\overline{\eta}}^{\mu\nu} \vec{\nabla}' 
\cdot \vec{\nabla} - \overline{\partial}^{\prime \mu} \overline{\partial}^{\nu}
\Bigr] \left[{ \frac18 \ln^2(H^2 {\Delta x}^2) + \frac12 \ln(H^2 {\Delta x}^2) 
\over \eta^2 \eta^{\prime 2}} \right] \right\} \; . \qquad \label{vacpol}
\end{eqnarray}
Note that it is completely integrable and gauge invariant, and that we have
everywhere taken $D=4$.

\section{Going on-shell}

Taking the vacuum polarization ``on-shell'' in a locally de Sitter background 
is complicated by the fact that the ``in'' vacuum is not equal to the ``out''
vacuum. This is obvious from the fact that there is particle creation. The 
photon wave function is therefore not the matrix element of $A_{\mu}(x)$ 
between a 1-photon in-state and the out-vacuum. It is rather the matrix
element of $A_{\mu}(x)$ between a 1-photon state and the Bunch-Davies vacuum, 
both prepared at $t=0$. The field equations obeyed by this matrix element are 
given by varying the Schwinger-Keldysh effective action 
\cite{Schwinger,Jordan}.

The rules for computing in the Schwinger-Keldysh formalism are simple. The
diagrams have the same topology as those of Feynman but the endpoints of 
lines bear either a ``$+$'' or a ``$-$'' polarity. All external lines are 
$+$, whereas vertices can be either all $+$ or all $-$. The $+$ vertices 
are the same as those of the standard Feynman rules; the $-$ vertices are 
conjugated. There are $++$, $+-$, $-+$ and $--$ propagators. All of them are 
the same function (\ref{prop}) of the appropriate version of the modified de
Sitter length function $y{\scriptscriptstyle \pm \pm}(x;x') \equiv a(\eta) 
a(\eta') H^2 {\Delta x}^2_{\scriptscriptstyle \pm \pm}$, where we define,
\begin{eqnarray}
{\Delta x}^2_{\scriptscriptstyle ++}(x;x') & \equiv & \Vert \vec{x} - \vec{x}' 
\Vert^2 - (\vert \eta - \eta' \vert - i \delta)^2 \; , \label{dx++} \\
{\Delta x}^2_{\scriptscriptstyle +-}(x;x') & \equiv & \Vert \vec{x} - \vec{x}' 
\Vert^2 - (\eta - \eta' + i \delta)^2 \; , \label{dx+-} \\
{\Delta x}^2_{\scriptscriptstyle -+}(x;x') & \equiv & \Vert \vec{x} - \vec{x}' 
\Vert^2 - (\eta - \eta' - i \delta)^2 \; , \label{dx-+} \\
{\Delta x}^2_{\scriptscriptstyle --}(x;x') & \equiv & \Vert \vec{x} - \vec{x}' 
\Vert^2 - (\vert \eta - \eta' \vert + i \delta)^2 \; . \label{dx--}
\end{eqnarray}
${i\Delta}_{\scriptscriptstyle ++}(x;x')$ and ${i\Delta}_{\scriptscriptstyle 
+-}(x;x')$ are equal for $\eta' > \eta$, hence the $++$ and $+-$ contributions 
cancel whenever $\eta' > \eta$. When the $x^{\mu}$ and $x^{\prime \mu}$ are 
spacelike related, the real part of $y(x;x')$ is positive; when they are 
timelike, the real part of $y(x;x')$ is negative. Therefore meromorphic
functions of $y_{\scriptscriptstyle \pm \pm}(x;x')$ agree for spacelike
separation for $\delta \rightarrow 0$. That is why the $++$ and $+-$ 
contributions cancel when $x^{\mu\prime}$ strays outside the past lightcone of 
$x^{\mu}$. Inside the past lightcone the $++$ and $+-$ propagators are 
conjugate.

The Schwinger-Keldysh effective action involves background fields  
$A^+_{\mu}(x)$ for the $+$ lines and $A^-_{\mu}(x)$ for the $-$ lines,
\begin{eqnarray}
\lefteqn{\Gamma[A^+;A^-] = -\frac14 \int d^4x \Bigl\{ F^+_{\mu \nu} F^+_{\rho
\sigma} - F^-_{\mu \nu} F^-_{\rho \sigma} \Bigr\} g^{\mu \rho} g^{\nu \sigma}
\sqrt{-g} } \nonumber \\
& & + \frac12 \int d^4x \, d^4x' \Biggl\{ A^+_{\mu}(x) \Bigl[\mbox{}^{\mu}
\Pi^{\nu}_{\scriptscriptstyle ++}\Bigr](x;x') A^+_{\nu}(x') + A^+_{\mu}(x) 
\Bigl[\mbox{}^{\mu}\Pi^{\nu}_{\scriptscriptstyle +-}\Bigr](x;x') A^-_{\nu}(x') 
\nonumber \\
& & + A^-_{\mu}(x) \Bigl[\mbox{}^{\mu}\Pi^{\nu}_{\scriptscriptstyle -+}\Bigr](
x;x') A^+_{\nu}(x') + A^-_{\mu}(x) \Bigl[\mbox{}^{\mu}\Pi^{\nu}_{
\scriptscriptstyle --}\Bigr](x;x') A^-_{\nu}(x')\Biggr\} + O(A^4) . \quad \;\;
\end{eqnarray}
The various $\pm$ permutations of the vacuum polarization are all the same 
function (\ref{vacpol}) with the appropriate $\pm$ permutation 
(\ref{dx++}-\ref{dx--}) substituted for ``${\Delta x}^2$''.\footnote{This may
seem surprising because there are no $+-$ or $-+$ seagull graphs or 
counterterms, so the only diagram topology contributing to $\Bigl[\mbox{}^{
\mu}\Pi^{\nu}_{\scriptscriptstyle +-} \Bigr](x;x')$ or its conjugate is Fig.~2.
The apparent paradox is resolved by noting that the mixed permutations also
fail to produce local terms coming from partial integration. Owing to the 
absence of the temporal absolute value in ${\Delta x}^2_{\scriptscriptstyle 
+-}$ we have,
\begin{eqnarray}
{1 \over {\Delta x}^{4 - 2 \epsilon}_{\scriptscriptstyle +-}} = - {\partial^2 
\over 2 \epsilon (1 - \epsilon)} \left[{1 \over {\Delta x}^{2 - 2 \epsilon}_{
\scriptscriptstyle +-}} - {\mu^{- \epsilon} \over {\Delta x}^{2 - \epsilon}_{
\scriptscriptstyle +-}} \right] \longrightarrow - {\partial^2 \over 4} \left[ 
{\ln(\mu^2 {\Delta x}^2_{\scriptscriptstyle +-}) \over {\Delta x}^2_{
\scriptscriptstyle +-}} \right] \; . \nonumber
\end{eqnarray}}
The Schwinger-Keldysh field equations are obtained by varying the effective
action with respect to either background and then equating the $+$ and $-$
fields,
\begin{eqnarray}
\lefteqn{ {\delta \Gamma[A^+;A^-] \over \delta A^+_{\mu}(x)} \Biggl\vert_{
A^{\pm}_{\mu} = A_{\mu}} = \partial_{\nu} \Bigl( \sqrt{-g} g^{\nu \rho} 
g^{\mu \sigma} F_{\rho \sigma} \Bigr) } \nonumber \\
& & + \int d^4x' \left\{ \Bigl[\mbox{}^{\mu}\Pi^{\nu}_{\scriptscriptstyle ++}
\Bigr](x;x') + \Bigl[\mbox{}^{\mu}\Pi^{\nu}_{\scriptscriptstyle +-}\Bigr](x;x')
\right\} A_{\nu}(x') + O(A^3) = 0 \; . \qquad \label{SKeqn}
\end{eqnarray}
Note that we have exploited the relation $\Bigl[\mbox{}^{\nu}\Pi^{\mu}_{
\scriptscriptstyle -+}\Bigr](x';x) = \Bigl[\mbox{}^{\mu}\Pi^{\nu}_{
\scriptscriptstyle +-}\Bigr](x;x')$.

Because electromagnetism is conformally invariant for $D=4$ the order $\hbar^0$
field equations are the same (in conformal coordinates) as those of flat space.
The general classical solution is therefore a superposition of plane waves,
\be
A_{\mu}^0(x) = \epsilon_{\mu}(\vec{k}) e^{i k \cdot x} \; ,
\ee
where $k^0 = \Vert \vec{k} \Vert$ and the Lorentz gauge polarization
vectors obey $\epsilon_0 = 0 = k \cdot \epsilon$. As in flat space we check
for quantum corrections by expanding the solution in powers of $\hbar$,
\be
A_{\mu}(x) = A^0_{\mu}(x) + A^1_{\mu}(x) + \dots \; .
\ee
and then segregating all terms of the same order in the field equations. 
Potential one loop corrections are determined by the equation,
\be
\Bigl[ \partial^2 \eta^{\mu \nu} - \partial^{\mu} \partial^{\nu} \Bigl]
A^1_{\nu}(x) = - \int d^4x' \left\{ \Bigl[\mbox{}^{\mu}\Pi^{\nu}_{
\scriptscriptstyle ++}\Bigr](x;x') + \Bigl[\mbox{}^{\mu}\Pi^{\nu}_{
\scriptscriptstyle +-}\Bigr](x;x') \right\} A^0_{\nu}(x') \; . \label{theeqn}
\ee
The temporal integration in this equation begins at the initial
time of $\eta' = -H^{-1}$. Its upper limit is irrelevant as long as it comes
later that $\eta$ because the cancellation between $++$ and $+-$ contributions 
eliminates contributions from any point $x^{\prime \mu}$ which is outside
the past lightcone of $x^{\mu}$.

From (\ref{vacpol}) we see that the vacuum polarization can be expressed as
the sum of four distinct terms,
\begin{eqnarray}
\lefteqn{\Bigl[\mbox{}^{\mu}\Pi^{\nu}_{\scriptscriptstyle \pm\pm}\Bigr](x;x') = 
{i e^2 \over 8 \pi^4} \left\{ \Bigl[ \eta^{\mu \nu} \partial' \cdot \partial - 
\partial^{\prime \mu} \partial^{\nu} \Bigr] \left[ F({\Delta x}^2_{\pm\pm}) +
{G({\Delta x}_{\pm\pm}) \over \eta \eta'} \right. \right. } \nonumber \\
& & \left. \left. \mp {i \pi^2 \over 3} \ln(a) \delta_{\pm \pm} 
\delta^4(x - x') \right] - \Bigl[ {\overline{\eta}}^{\mu\nu} \vec{\nabla}' 
\cdot \vec{\nabla} - \overline{\partial}^{\prime \mu} \overline{\partial}^{\nu}
\Bigr] {H({\Delta x}^2_{\pm\pm}) \over \eta^2 \eta^{\prime 2}} \right\} \; . 
\label{newpol}
\end{eqnarray}
Note that the conformal anomaly only contributes for the $++$ and $--$ cases.
The functions of ${\Delta x}^2_{\pm\pm}$ in the other three terms are,
\begin{eqnarray}
F({\Delta x}^2) & \equiv & \partial^2 \left[ {\ln(\mu^2 {\Delta x}^2) \over
24 {\Delta x}^2} \right] = \partial^4 \left[ \frac1{192} \ln^2(\mu^2 {\Delta
x}^2) - \frac1{96} \ln(\mu^2 {\Delta x}^2) \right] \; , \\
G({\Delta x}^2) & \equiv & {\frac12 \ln(H^2 {\Delta x}^2) + 1 \over
{\Delta x}^2} = \partial^2 \left[ \frac1{16} \ln^2(H^2 {\Delta x}^2) + \frac18 
\ln(H^2 {\Delta x}^2) \right] \; , \qquad \label{Gdef} \\
K({\Delta x}^2) & \equiv & \frac18 \ln^2(H^2 {\Delta x}^2) + \frac12 \ln(H^2 
{\Delta x}^2) \; .
\end{eqnarray}
It is useful to make the additional definitions, $F({\Delta x}^2) \equiv
\partial^4 f({\Delta x}^2)$ and $G({\Delta x}^2) \equiv \partial^2 g({\Delta
x}^2)$.

Of course it is easy to compute the conformal anomaly's contribution to the
right hand side of (\ref{theeqn}). Since this term is local we can actually
derive it for an arbitrary vector potential and then specialize to the
classical solution,
\begin{eqnarray}
C^{\mu}(x) & \equiv & - {e^2 \over 24 \pi^2} \int d^4x' \Biggl\{ 
\Bigl[ \eta^{\mu \nu} \partial' \cdot \partial - \partial^{\prime \mu} 
\partial^{\nu} \Bigr] \ln(a) \delta^4(x - x') \Biggr\} A_{\nu}(x') , \qquad \\
& = & {e^2 \over 24 \pi^2} \partial_{\nu} \Bigl( \ln(a) F^{\nu \mu}(x)
\Bigr) \; , 
 \label{anomaly1} \\
& \rightarrow & {i e^2 \over 24 \pi^2} H k a(\eta) \epsilon^{\mu}(\vec{k})
e^{i k \cdot x} \; .
\end{eqnarray}
$C^{\mu}(x)$ completely dominates the classical term by virtue of the factor of 
$a(\eta)$, but it is much less significant than the $a^2(\eta)$ associated
with a true photon mass. Interestingly, the conformal contribution is also 
purely dispersive.

Since the anomaly contribution~(\ref{anomaly1}) is local,  
it is instructive to combine it with the classical action $S_0$
into an effective action,
\begin{eqnarray}
 S_0 + \delta S_{\rm anom} &=& -\frac 12\int d^4 x d^4 x'
 A_\mu(x) 
 \bigg\{(\eta^{\mu\nu}\partial'\cdot\, \partial - {\partial^\mu}'\partial^\nu) 
\nonumber\\
 &\times& \bigg[\bigg(1+\frac{e^2}{24\pi^2}\ln\Big(\frac{a}{a_0}\Big)\bigg) 
\delta^4(x-x')\bigg]\bigg\} A_\nu(x') .
\label{anomaly-action}
\end{eqnarray}
More generally, for a nonabelian gauge theory $G$ coupled to $N_f$ Dirac
fermions and $N_s$ complex scalars in representation $r$, we 
have~\cite{SQED,Dolgov,Prokopec:2001}
\begin{eqnarray}
&& S_0 + \delta S_{\rm anom} = -\frac 14\int d^4 x
\nonumber\\
&\times& 
   \bigg[1 + \frac{\alpha_g}{3\pi}\Big(N_sC(r)
              + 4N_fC(r) 
  - 11 C_2(G)\Big) \ln\Big(\frac{a}{a_0}\Big)\bigg]
       \eta^{\mu\rho}\eta^{\nu\sigma}F_{\mu\nu} F_{\rho\sigma},
\quad
\label{anomaly-action2}
\end{eqnarray}
where $g$ is the gauge coupling constant, $\alpha_g = g^2/4\pi$, $C_2(G)$ is 
the quadratic Casimir of the adjoint representation of $G$,
which can be defined in terms of the generators of the adjoint
representation $T^a$ and the group structure constants 
$f^{acd}$ as ${\rm tr} [T^aT^b] = -f^{acd}f^{bcd} = -C_2(G)\delta ^{ab}$,
and $C(r)$ is defined in terms of the generators $t^a(r)$ of 
representation $r$ as ${\rm tr}  [t^a(r)t^b(r)] = C(r)\delta^{ab}$.
For example, for SU(N), $C_2 = N$, and for fermions and scalars in 
the fundamental representation of $SU(N)$, $C(r={\rm fundamental}) = 1/2$. 
The effect of the anomaly~(\ref{anomaly-action2}) on the photon dynamics
is quite moderate~\cite{Prokopec:2001}.
Since the photon mass contribution is parametrically much larger, 
we expect the photon dynamics to be affected much more by the 
photon mass~\cite{DDPT1}.

Our strategy for evaluating the other three contributions is to exploit the
fact that the range of $x^{\prime \mu}$ is independent of $x^{\mu}$ to first
pull the derivatives outside the integration. We then combine the $++$ and
$+-$ terms to obtain an integrand which is only nonzero for $x^{\prime \mu}$
within the past lightcone of $x^{\mu}$. The final step is a sometimes lengthy
series of asymptotic expansions under the assumption that the mode under
study went super-horizon long ago (hence $k \ll H a(\eta) = -1/\eta$) after a
long period of inflation (hence $H \ll k$). We shall organize these 
expansions in terms of two dimensionless parameters,
\be
y \equiv - k \eta \qquad {\rm and} \qquad w \equiv \frac{k}{H} 
 \label{scales}
\; .
\ee
The physically interesting region is $0 < y \ll 1 \ll w \ll a$. At horizon
crossing one would have $y \approx 1$ and $w \approx a$. Evolution of the 
physical scales in de Sitter inflation is illustrated in figure~4. The 
physical wave length $\lambda_{\rm phys} \equiv 2\pi a/k = 2\pi/Hy$ grows 
with time, while the Hubble radius remains constant.
\begin{figure}[htbp]
\centerline{\hspace{.0in} 
\epsfig{file=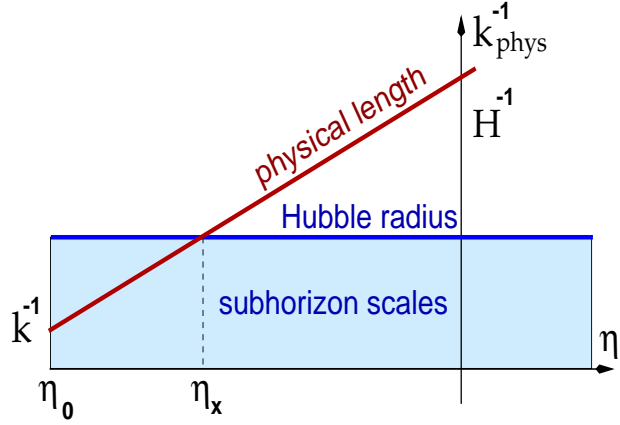, width=3.2in,height=2.2in}
}
\vskip -0.1in
\label{figure-two}
\caption[fig2]{
  \small Evolution of the physical scales in de Sitter inflation. Horizon
         crossing occurs at $\eta_x$. }
\end{figure}

Although it is the least important of the four terms, an exact result can be
obtained for the ``F'' contribution to the right hand side of (\ref{theeqn}).
We begin by reflecting the $x^{\prime \mu}$ derivatives ($\partial_{\mu}^{
\prime} F({\Delta x}^2) = - \partial_{\mu} F({\Delta x}^2)$) and then pulling
all derivatives outside the integration,
\begin{eqnarray}
\lefteqn{F^{\mu}(x) } \nonumber \\
& & \equiv - {i e^2 \over 8 \pi^4} \int d^4x' \Biggl\{ 
\Bigl[ \eta^{\mu \nu} \partial' \cdot \partial - \partial^{\prime \mu} 
\partial^{\nu} \Bigr] \Bigl[F({\Delta x}^2_{++}) - F({\Delta x}^2_{+-})
\Bigr] \Biggr\} \epsilon_{\nu} e^{i k \cdot x'} , \qquad \\
& & = {i e^2 \over 8 \pi^4} \Bigl[\epsilon^{\mu} \partial^2 - \epsilon 
\cdot \partial \partial^{\mu} \Bigr] \partial^4 \int d^4x' \Bigl\{ f({\Delta
x}^2_{++}) - f({\Delta x}^2_{+-}) \Bigr\} e^{i k \cdot x'} \; . 
\end{eqnarray}
The difference between the $++$ and $+-$ contributions vanishes for $x^{\prime 
\mu}$ outside the past lightcone of $x^{\mu}$,
\be
f({\Delta x}^2_{++}) - f({\Delta x}_{+-}) = {i \pi \over 48} \Bigl\{ \ln\Bigl[
\mu^2 ( {\Delta \eta}^2 - \Vert \vec{x} \Vert^2 ) \Bigr] - 1 \Bigl\} 
\theta({\Delta \eta}) \theta({\Delta \eta} - \Vert \vec{x} \Vert ) \; .
\ee
We next change variables to ${\Delta x}^{\mu} \equiv x^{\mu} - x^{\prime \mu}$
and perform the angular integrations,
\begin{eqnarray}
\lefteqn{F^{\mu}(x) = {i e^2 \over 8 \pi^4} \Bigl[\epsilon^{\mu} 
\partial^2 - \epsilon \cdot \partial \partial^{\mu} \Bigr] \partial^4 e^{i k 
\cdot x} \int_0^{\eta + {\scriptscriptstyle H^{-1}}} d{\Delta \eta} e^{i k 
{\Delta \eta}} } \nonumber \\
& & \hspace{2cm} \times 4 \pi \int_0^{\Delta \eta} dr r^2 {\sin(k r) \over kr}
{i \pi \over 48} \Bigl\{ \ln\Bigl[ \mu^2 ( {\Delta \eta}^2 - r^2) \Bigr] 
- 1 \Bigr\} \; .
\end{eqnarray}

The radial integration involves a combination of special functions 
$\xi(k\Delta\eta)$ that we will meet again,
\begin{eqnarray}
\lefteqn{\int_0^{\Delta \eta} dr r \sin(k r) \Bigl\{ \ln\Bigl[ \mu^2 ({\Delta 
\eta}^2 - r^2) \Bigr] - 1 \Bigr\} } \nonumber \\
& & = {\Delta \eta}^2 \int_0^1 dx x \sin(k {\Delta \eta} x) \Bigl\{ 2 \ln(\mu 
{\Delta \eta}) - 1 + \ln(1 - x^2) \Bigr\} \; , \\
& & = \frac1{k^2} \Bigl[ \sin(k {\Delta \eta}) - k {\Delta \eta} \cos(k
{\Delta \eta}) \Bigr] \Bigl[ 2 \ln(\mu {\Delta \eta}) - 1 \Bigr] + 
{\Delta \eta}^2 \xi(k {\Delta \eta}) \; .
\end{eqnarray}
Our definition for $\xi$ is,\footnote{We use the following notation for
the sine and cosine integrals:
\begin{eqnarray}
{\rm si}(x) & \equiv & - \int_x^{\infty} dt {\sin(t) \over t} = -\frac{\pi}2
+ \int_0^x dt {\sin(t) \over t} \; , \nonumber \\
{\rm ci}(x) & \equiv & -\int_x^{\infty} dt {\cos(t) \over t} = \gamma + \ln(x)
+ \int_0^x dt {\cos(t) -1 \over t} \; . \nonumber 
\end{eqnarray}}
\begin{eqnarray}
\xi(\alpha) & \equiv & \int_0^1 dx x \sin(\alpha x) \ln(1 - x^2) \; , \\
& = & \frac2{\alpha^2} \sin(\alpha) - \frac1{\alpha^2} [\cos(\alpha) +
\alpha \sin(\alpha)] \left[ {\rm si}(2 \alpha) + \frac{\pi}2 \right] 
\nonumber \\
& & \hspace{2cm} + \frac1{\alpha^2} [\sin(\alpha) - \alpha \cos(\alpha)]
\left[ {\rm ci}(2 \alpha) - \gamma - \ln\Bigl(\frac{\alpha}2\Bigr) \right] \; .
\end{eqnarray}
Here $\gamma \approx .577$ is Euler's constant.
The following small and large $\alpha$ expansions are sometimes useful,
\begin{eqnarray} 
\lefteqn{\xi(\alpha) = -\left[ \frac89 - \frac23 \ln(2)\right] \alpha + 
O(\alpha^3) \; , } \\
& & = {\ln(\frac{\alpha}2) \over \alpha} \cos(\alpha) + \frac1{\alpha} \left[ 
\gamma \cos(\alpha) - \frac{\pi}2 \sin(\alpha) \right] + O\left({\ln( \alpha) 
\over \alpha^2}\right) \; . \qquad
\end{eqnarray}

Combining the previous results gives $F^{\mu}(x) = \Bigl[ \epsilon^{\mu}
\partial^2 - \epsilon \cdot \partial \partial^{\mu} \Bigr] \partial^4 
e^{k \cdot x} I(\eta)$ where,
\begin{eqnarray}
\lefteqn{I(\eta) \equiv - {e^2 \over 96 \pi^2 k^3} \int_0^{\eta + 
{\scriptscriptstyle H^{-1}}} d{\Delta \eta} e^{i k {\Delta \eta}}} \nonumber \\
& & \times \left\{ \Bigl[ \sin(k {\Delta \eta}) - k {\Delta \eta} \cos(k
{\Delta \eta}) \Bigr] \Bigl[ 2 \ln(\mu {\Delta \eta}) - 1 \Bigr] + 
k^2 {\Delta \eta}^2 \xi(k {\Delta \eta}) \right\} . \qquad 
\end{eqnarray}
The only $\vec{x}$ dependence in this expression resides in the outer 
exponential factor. We can act with the spatial derivatives and then
commute the temporal derivatives through the exponential to obtain,
\begin{eqnarray}
\Bigl[\epsilon^{\mu} \partial^2 - \epsilon \cdot \partial \partial^{\mu} 
\Bigr] \partial^4 e^{i k \cdot x} I(\eta) & = & \epsilon^{\mu} (-\partial_0^2
- k^2)^3 e^{i k \cdot x} I(\eta) \; , \\
& = & - \epsilon^{\mu} e^{i k \cdot x} (\partial_0^2 - 2 i k \partial_0)^3
I(\eta) \; .
\end{eqnarray}
After commutation the only $\eta$ dependence to the right of the derivatives 
resides in the upper limit of the $\Delta \eta$ integration. So one of the 
$\partial_0$'s acts to undo the integral and we are left with the messy task 
of taking the remaining derivatives. The final result is,
\be
F^{\mu}(\eta) = -{e^2 \over 24 \pi^2} {k H \over 1 - a^{-1}} \left\{i + {H
\over k} {\cos[\frac{k}{H} ( 1 - a^{-1})] \over 1 - a^{-1}} e^{i \frac{k}{H}
(1 - a^{-1})} \right\} \epsilon^{\mu} e^{i k \cdot x} \; . \label{Ffin}
\ee

Because it lacks positive powers of the scale factor, $F^{\mu}(x)$ is
far weaker than $C^{\mu}(x)$. In fact it is only nonzero because the
process begins at $t=0$ ($\eta = -H^{-1}$), rather than at $t = -\infty$ ($\eta
= -\infty$). This is as it should be since the ``F'' contributions to the 
vacuum polarization are the same as those of flat space, which vanish when 
exactly on-shell. We get a nonzero result because beginning at a finite time 
precludes one from going precisely on-shell. This is an important check on 
the accuracy and consistency of the process.

The ``K'' contributions are reduced in almost the same way. They have zero 
temporal component by definition. Their spatial components are,
\begin{eqnarray}
\lefteqn{K^i(x) } \nonumber \\
& & \equiv {i e^2 \over 8 \pi^4} \int d^4x' \Biggl\{ 
\Bigl[ \delta^{ij} \vec{\nabla}' \cdot \vec{\nabla} - \partial^{\prime i} 
\partial^{j} \Bigr] \left[{H({\Delta x}^2_{++}) - H({\Delta x}^2_{+-}) \over
\eta^2 \eta^{\prime 2} } \right] \Biggr\} \epsilon_j e^{i k \cdot x'} , 
\qquad \\
& & = - {i e^2 \over 8 \pi^4 \eta^2} \Bigl[\epsilon^i \nabla^2 - \vec{\epsilon}
\cdot \vec{\nabla} \partial^i \Bigr] \int d^4x' {e^{ik \cdot x'} \over \eta^{
\prime 2}} \Bigl\{ H({\Delta x}^2_{++}) - H({\Delta x}^2_{+-}) \Bigr\} \; , \\
& & = -{i e^2 H^2 \over 8 \pi^4} a^2 \Bigl[\epsilon^i \nabla^2 - \vec{\epsilon}
\cdot \vec{\nabla} \partial^i \Bigr] e^{i k \cdot x} \int_0^{\eta + 
{\scriptscriptstyle H^{-1}}} d{\Delta \eta} {e^{i k {\Delta \eta}} \over 
({\Delta \eta} - \eta)^2 } \nonumber \\
& & \hspace{2cm} \times 4 \pi \int_0^{\Delta \eta} dr r^2 {\sin(k r) \over kr}
{i \pi \over 2} \Bigl\{ \ln\Bigl[ H^2 ( {\Delta \eta}^2 - r^2) \Bigr] + 2
\Bigr\} \; , \\
& & = {e^2 H^2 \over 4 \pi^2 k^3} a^2 \Bigl[\epsilon^i \nabla^2 - \vec{\epsilon}
\cdot \vec{\nabla} \partial^i \Bigr] e^{i k \cdot x} \int_0^{\eta +
{\scriptscriptstyle H^{-1}}} d{\Delta \eta} {e^{i k {\Delta \eta}} \over 
({\Delta \eta} - \eta)^2} \nonumber \\
& & \times \Biggl\{ \Bigl[\sin(k {\Delta \eta}) - k {\Delta \eta} 
\cos(k {\Delta \eta}) \Bigr] \Bigl[ 2 \ln(H {\Delta \eta}) + 2 \Bigr] + 
k^2 {\Delta \eta}^2 \xi(k {\Delta \eta}) \Biggr\} \; , \\
& & = -{e^2 H^2 \over 4 \pi^2} a^2 \epsilon^i e^{i k \cdot x} \int_0^{w-y} dz 
{e^{iz} \over (z + y)^2} \nonumber \\
& & \hspace{2cm} \times \left\{ [\sin(z) - z \cos(z)] \Bigl[2 \ln(z) - 2 
\ln(w) + 2 \Bigr] + z^2 \xi(z) \right\} \; .
\end{eqnarray}
(Recall that we define $y \equiv -k \eta \ll 1$ and $w \equiv k/H \gg 1$.) 
The next step would be making an asymptotic expansion of the integral but 
this would be wasted effort because almost the same integral occurs with the 
opposite sign in the ``G'' contributions.

The contributions from $G$ (\ref{Gdef}) to the right hand side of 
(\ref{theeqn}) are the most difficult to evaluate owing to the combination 
of temporal derivatives and the factor of $1/\eta \eta'$. Because of this it 
is desirable to act with some of the derivatives and exploit the 
transversality of the polarization vector at an earlier stage than with the 
other terms,
\begin{eqnarray}
\lefteqn{G^{\mu}(x) } \nonumber \\
& & \equiv - {i e^2 \over 8 \pi^4} \int d^4x' \Bigl\{ \Bigl[ \eta^{\mu \nu} 
\partial' \cdot \partial - \partial^{\prime \mu} \partial^{\nu} \Bigr] 
\Bigl[{G({\Delta x}^2_{++}) - G({\Delta x}^2_{+-}) \over \eta \eta' } \Bigr] 
\Bigr\} \epsilon_{\nu} e^{i k \cdot x'} , \qquad \\
& & = - {i e^2 \over 8 \pi^4} \Bigl[\epsilon^{\mu} \partial^{\nu} - \eta^{\mu
\nu} \epsilon \cdot \partial \Bigr] \frac1{\eta} \partial^2 e^{i k \cdot x}
\nonumber \\
& & \hspace{2cm} \times \int d^4x' e^{-i k \cdot {\Delta x}} \partial_{\nu}' 
\left[{g({\Delta x}^2_{++}) - g({\Delta x}^2_{+-}) \over \eta' } \right] 
\; , \\
& & = {i e^2 \over 8 \pi^4} H^2 a^2 \epsilon^{\mu} e^{i k \cdot x} \Bigl[
\delta^{\nu}_{~ 0} + \eta (- \delta^{\nu}_{~ 0} \partial_0 + i k^{\nu}) \Bigr]
(\partial_0 - 2 i k) \partial_0 \nonumber \\
& & \hspace{2cm} \times \int d^4x' e^{-i k \cdot {\Delta x}} \partial_{\nu}' 
\left[{g({\Delta x}^2_{ ++}) - g({\Delta x}^2_{+-}) \over \eta' } \right] \; .
\label{actout}
\end{eqnarray}
At this point it is best to partially integrate the $\partial_{\nu}'$. Owing to
the cancellation between $g({\Delta x}^2_{++})$ and $g({\Delta x}^2_{+-})$ 
outside the past light cone, only the lower, temporal surface term can survive,
\begin{eqnarray}
\lefteqn{\int d^4x' e^{-i k \cdot {\Delta x}} \partial_{\nu}' \left[{
g({\Delta x}^2_{ ++}) - g({\Delta x}^2_{+-}) \over \eta' } \right] } 
\nonumber \\
& & = -i k_{\nu} \int d^4x' {e^{-i k \cdot {\Delta x}} \over \eta'} \Bigl[
g({\Delta x}^2_{ ++}) - g({\Delta x}^2_{+-}) \Bigr] \nonumber \\
& & \hspace{2cm} + \delta^0_{~ \nu} H \int d^3x' e^{-i k \cdot {\Delta x}} 
\Bigl[ g({\Delta x}^2_{ ++}) - g({\Delta x}^2_{+-}) \Bigr] \Biggl\vert_{\eta' 
= - H^{-1}} . \qquad \label{parts}
\end{eqnarray}
Since $k^{\nu} k_{\nu} = 0$ we see that only the $\nu = 0$ component survives.

The spatial integral can be evaluated by familiar techniques,
\begin{eqnarray}
\lefteqn{\int d^3x' e^{-i k \cdot {\Delta x}} \Bigl[ g({\Delta x}^2_{ ++}) - 
g({\Delta x}^2_{+-}) \Bigr] } \nonumber \\
& & = e^{i k {\Delta \eta}} 4 \pi \int_0^{{\Delta \eta}} dr r^2 {\sin(k r) 
\over kr} {i \pi \over 4} \Bigl\{ \ln\Bigl[ H^2 ({\Delta \eta}^2 - r^2)\Bigr]
+ 1 \Bigr\} \; , \\
& & = {\pi^2 i \over k^3} e^{i k {\Delta \eta}} \Biggl\{ \Bigl[\sin(k {\Delta 
\eta}) - k {\Delta \eta} \cos(k {\Delta \eta}) \Bigr] \nonumber \\
& & \hspace{3cm} \times \Bigl[ 2 \ln(k {\Delta \eta}) -2 \ln(w) + 1 \Bigr] + 
k^2 {\Delta \eta}^2 \xi(k {\Delta \eta}) \Biggr\} \; , \qquad \\
& & \equiv {\pi^2 i \over k^3} \Xi(k {\Delta \eta},w) \; .
\end{eqnarray}
The function $\Xi(x,w)$ has the following asymptotic expansions for small
and large $x$ respectively,
\begin{eqnarray}
&&\!\!\Xi(x,w) =  \frac23 x^3 
  \biggl[\ln\Bigl(\frac{2 x}{w}\Bigr) - \frac56 \biggr] + 
O\Bigl(x^5 \ln(x)\Bigr) \; , \\
& = &\!\! x e^{i x} \Bigl[\Bigl(2 \ln(w) - \ln(2 x) - (1-\gamma)\Bigr)
\cos(x) - \frac{\pi}2 \sin(x)\Bigr]  
+ O\Bigl(\ln(xw) \Bigr) . \qquad\;
\end{eqnarray}

Substituting the spatial integral into (\ref{parts}), and inserting the result 
into (\ref{actout}) gives,
\begin{eqnarray}
\lefteqn{G^{\mu}(x) = -{e^2 H^2 \over 8 \pi^2} a^2 \epsilon^{\mu} 
e^{i k \cdot x} (1 - y \partial_y) (2 - i \partial_y) \partial_y \int_0^{w-y}
dz {\Xi(z,w) \over z + y} } \nonumber \\
& & \hspace{1cm} -{i e^2 H^2 \over 8 \pi^2} a^2 \epsilon^{\mu} e^{i k \cdot x} 
(1 - y \partial_y - i y) (2 - i \partial_y) \partial_y {\Xi(w - y,w) \over w}
\; , \\
& & = {e^2 H^2 \over 8 \pi^2} a^2 \epsilon^{\mu} e^{i k \cdot x} \Bigl(2 -
i \partial_y - 2 y \partial_y + i y \partial_y^2\Bigr) \int_0^{w - y} dz 
{\Xi(z,w) \over (z + y)^2} \nonumber \\
& & \hspace{1cm} + {e^2 H^2 \over 8 \pi^2 } a^2 \epsilon^{\mu} e^{i k \cdot x} 
\Bigl(1 - i \partial_y -2 y \partial_y + i y \partial_y^2\Bigr) (2 - i 
\partial_y) {\Xi(w - y,w) \over w } . \qquad \label{twolines}
\end{eqnarray}
(Recall that we define $y \equiv -k \eta \ll 1$ and $w \equiv k/H \gg 1$.)
Now note that the first term on the top line almost exactly cancels 
$K^{\mu}(x)$,
\begin{eqnarray}
\lefteqn{K^{\mu}(x) + {e^2 H^2 \over 8 \pi^2} a^2 \epsilon^{\mu} e^{i k
\cdot x} \times 2 \int_0^{w - y} dz {\Xi(z,w) \over (z + y)^2} } \nonumber \\
& & = -{e^2 H^2 \over 4 \pi^2} a^2 \epsilon^{\mu} e^{i k \cdot x} \int_0^{w 
- y} dz {e^{i z} \over (z + y)^2} \Bigl[ \sin(z) - z \cos(z) \Bigr] \; , \\
& & = - {e^2 H^2 \over 4 \pi^2} a^2 \epsilon^{\mu} e^{i k \cdot x} \left\{
\int_0^{w} dz {e^{i z} \over z^2} \Bigl[ \sin(z) - z \cos(z) \Bigr] + O(y) 
\right\} \; , \\
& & = {e^2 H^2 \over 8 \pi^2} a^2 \epsilon^{\mu} e^{i k \cdot x} \Bigl\{
\ln(w) + O(1) \Bigr\} \; .
\end{eqnarray}
This turns out to be the magnitude of the leading order contribution for
super-horizon modes late during inflation.

Another leading order contribution comes from the second term on the top line
of (\ref{twolines}). To save space we suppress the prefactor of ${e^2 H^2
\over 8 \pi^2} a^2 \epsilon^{\mu} e^{i k \cdot x}$,
\be
-i \partial_y \int_0^{w-y} dz {\Xi(z,w) \over (z + y)^2} = \frac{i}{w^2}
\Xi(w-y,w) + 2 i \int_0^{w-y} dz { \Xi(z,w) \over (z + y)^3} \; .
\ee
The first term on the left hand side is of order $\ln(w)/w$ but the integral
can make a leading order contribution. Since the integrand converges for $y=0$
and $w \rightarrow \infty$, the desired leading term comes from the explicit 
factor of $\ln(w)$ in the function $\Xi(z,w)$,
\begin{eqnarray}
\lefteqn{2 i \int_0^{w - y} dz {\Xi(z,w) \over (z + y)^3}} \nonumber \\
& & = - 4 i \ln(w) \int_0^{\infty} dz {e^{iz} \over z^3} \Bigl[ \sin(z) - z 
\cos(z) \Bigr] + O(1) \; , \\
& & = -4 i \ln(w) \times {i \over 4 z^2} \Bigl[e^{2 i z} - 1 - 2 i z \Bigr]
\Biggl\vert_0^{\infty} + O(1) \; , \\
& & = 2 \ln(w) + O(1) \; .
\end{eqnarray}

The final leading order contribution comes from the first term on the bottom 
line of (\ref{twolines}). It is useful to first note the exact identity,
\begin{eqnarray}
\lefteqn{ (2 + i \partial_x) \Xi(x,w) = } \nonumber \\
& & - x \Bigl\{ \Bigl[ {\rm ci(2x)} + \ln(2x) + 1 - \gamma - 2 \ln(w) \Bigr] 
+ i \Bigl[ {\rm si}(2 x) + \frac{\pi}2 \Bigr] \Bigr\} + e^{i x} \Bigl\{ 
\sin(x) \nonumber \\
& & \hspace{1cm} \times \Bigl[ {\rm ci(2x)} + \ln(2x) + 3 - \gamma - 2 \ln(w) 
\Bigr] - i \cos(x) \Bigl[ {\rm si}(2 x) + \frac{\pi}2 \Bigr] \Bigr\} . \qquad
\end{eqnarray}
Setting $x = w - y$ we conclude,
\be
(2 - i \partial_y) {\Xi(w - y, w) \over w} = \ln(w) + O(1) \; .
\ee
Combining all of the leading terms we obtain the following result for the 
right hand side of Eq.~(\ref{theeqn}),
\be
C^\mu(x) + F^\mu(x) + G^{\mu}(x) + K^{\mu}(x) 
  = a^2(\eta) {e^2 H^2 \over 2 \pi^2} 
\Bigl\{ \ln(w) + O(1) \Bigr\}
{A^0}^{\mu}(x)
 \; .
\label{rhs-of-82}
\ee
Note that the contributions 
from $C^{\mu}(x)$ and $F^{\mu}(x)$
are down by factors of $w/a \equiv k_{\rm phys}/H \ll 1$ 
and $(w/a)^2$, respectively.
Equation~(\ref{rhs-of-82}) is consistent with a photon mass of,
\be
 m^2_{\gamma} = {e^2 H^2 \over 2 \pi^2} \ln\Big(\frac kH\Big).
\ee
It is important to keep in mind that this result is perturbative. 
A full nonperturbative analysis of~(\ref{theeqn}) is thus required in order
to calculate reliably the photon mass. 

\section{Discussion}

We have presented a long calculation in perturbative quantum field theory and
it is worth commenting on the matter of reliability. Our result (\ref{vacpol})
for the vacuum polarization passes many important consistency checks. The 
first of these is gauge invariance. This is not trivial even in flat space. 
In a locally de Sitter background it requires a horrifying series of seemingly 
unrelated terms to combine into transverse projection operators. Yet they do.

Another important accuracy check is that there are no new ultraviolet 
divergences. There should not be if the theory is to stay renormalizable
because there are no new counterterms in de Sitter background. Related to 
this is the fact that our result has the correct flat space limit. A final
check is that the conformal anomaly term agrees with standard results
\cite{Dolgov,SQED}.

 The structure of the self energy~(\ref{vacpol})
is very similar to that of the photon self energy in thermal QED.
The self energy (in momentum space) is usually parametrized
by two functions,
$\Pi^{\mu\nu}(k) = P_T^{\mu\nu} \Pi_t(k) + P_L^{\mu\nu} \Pi_l(k)$,
where $P_T^{\mu\nu} = \eta^{\mu i}\eta^{\nu j}(\delta_{ij} - k_ik_j/\vec k^2)$
and $P_L^{\mu\nu} = \eta^{\mu\nu} - k^\mu k^\nu/k^2 - P_T^{\mu\nu}$ are 
the (spatially) transverse and `longitudinal' 
(in fact time-like transverse) projectors.
Our result~(\ref{vacpol}) can be viewed as the space-time generalization
of the transverse and `logitudinal' vacuum polarizations,
\begin{eqnarray} 
 \Pi_t(x,x') &=& \partial'\cdot\partial\, \Pi^{(1)}(x,x')
              + \nabla'\cdot\nabla\, \Pi^{(2)}(x,x')
\nonumber\\
  \Pi_l(x,x') &=& \partial'\cdot\partial\,\Pi^{(1)}(x,x') ,
\label{Pi-t,l}
\end{eqnarray} 
where $\Pi^{(1)}$ and $\Pi^{(2)}$ are the transverse and
 spatially transverse contributions to~(\ref{vacpol})
({\it cf.} also Eq.~(\ref{A-B})).
In thermal QED vacuum polarization modifies the photon dynamics.
In the static limit `longitudinal' photons are
Debye-screened by the fermionic plasma, with the Debye mass, 
$m_D = e T/\sqrt{3}$, while transverse photons are screened
only dynamically. At high momenta $k\gg T$, the `longitudinal' 
modes become unphysical, while the transverse photons propagate as 
massive particles with the thermally induced mass $m_T = e T/\sqrt{6}$.
Based on the above mentioned similarities, we expect that a more detailed
study of the vacuum polarization~(\ref{vacpol}) should reveal analogous
physical effects on the photon dynamics in inflation. 

We are very confident about our result (\ref{vacpol}) for
$\Bigl[\mbox{}^{\mu}\Pi^{\nu}\Bigr](x,x')$. It has to be admitted that the 
process of going on-shell is much less well checked. This is also the most 
complicated part of the calculation. One important point is that (\ref{Ffin}), 
the flat space contribution $F^{\mu}(x)$, vanishes in the limit that the 
initial time is taken to negative infinity. Although $F^{\mu}(x)$ makes the 
weakest of the various contributions, its reduction is quite similar to that 
of the crucial $G^{\mu}(x)$ and $K^{\mu}(x)$ contributions. So the fact that 
$F^{\mu}(x)$ obeys an important correspondence limit partially checks
them as well.

Our result is consistent with a photon mass of $m^2_{\gamma} = {e^2 H^2 \over 
2 \pi^2} \ln(k/H)$. Interestingly, this is precisely what follows from the
Hartree-Fock estimate~(\ref{Hartree-Fock})
if one replaces the
time dependent factor of $\ln(a)$ by its value at horizon crossing, $\ln(k/H)$.
It is premature to make too much of this coincidence. Although our 1-loop
vacuum polarization is exact, all the work of taking it on-shell really
demonstrates is that one loop corrections to the classical photon wave function
become non-perturbatively large. 

To actually solve for the photon wave function and show that it approaches that 
of a massive photon requires two extensions of the current work. First, we 
must establish control over higher loop corrections. If the one loop correction
becomes large then why does the two loop correction not give an even bigger
effect? There is a curious parallel between what we must do and the problem 
that Schwinger faced for $D=2$, massless QED in flat space \cite{JS}. Just as 
it was possible to use the one loop result in that context, so we believe it 
will prove to be in this case. 

To understand how this can be, note that we know quite a lot about the general 
structure of the vacuum polarization. As a consequence of gauge invariance, 
spatial translation invariance and spatial rotational invariance, it must 
have the form,
\begin{eqnarray}
\Bigl[\mbox{}^{\mu}\Pi^{\nu}\Bigr](x;x') & = & \Bigl[ \eta^{\mu \nu} \partial' 
\cdot \partial - \partial^{\prime \mu} \partial^{\nu} \Bigr] A\Bigl(\eta,
\eta';{\Delta x}^2\Bigr) \nonumber \\
& & \hspace{2cm} + \Bigl[ {\overline{\eta}}^{\mu\nu} \vec{\nabla}' \cdot 
\vec{\nabla} - \overline{\partial}^{\prime \mu} \overline{\partial}^{\nu}\Bigr]
B\Bigl(\eta,\eta';{\Delta x}^2\Bigr) \; .
 \label{A-B}
\end{eqnarray}
The functions $A\Bigl(\eta,\eta';{\Delta x}^2\Bigr)$ and $B\Bigl(\eta,\eta';
{\Delta x}^2\Bigr)$ must be symmetric under interchange of $\eta$ and $\eta'$,
and they must have the dimensions of inverse length to the fourth power.
The full flat space result must reside in $A\Bigl(\eta,\eta';{\Delta x}^2
\Bigr)$, but it will be negligible in de Sitter background. Terms that matter
for de Sitter are those with factors of $1/\eta$ and $1/\eta'$. We believe 
quite strongly that $A\Bigl(\eta,\eta';{\Delta x}^2\Bigr)$ can contain at most 
$1/(\eta \eta')$ and that $B\Bigl(\eta,\eta';{\Delta x}^2\Bigr)$ can contain 
at most $1/(\eta^2 \eta^{\prime 2})$. 

Obviously each loop will contribute a factor of $e^2$. The only really 
difficult thing to guess is the number of logarithms. We believe --- though 
less strongly --- that the general result at $\ell$ loops is $A\Bigl(\eta,
\eta';{\Delta x}^2\Bigr)$ contains up to $\ell$ logarithms whereas $B\Bigl(
\eta,\eta';{\Delta x}^2\Bigr)$ contains up to $\ell+1$. We further believe
that in each case these logarithms can translate, after going on-shell, into
up to one factor of $\ln(k/H)$ for each loop order. Since there is also a
factor of $\alpha \equiv e^2/4\pi$ for each extra loop, 
this suggests that retaining only 
the one loop part would be reliable for modes which obey $\alpha 
\ln(k/H) < 1$. That is, the one loop term would dominate the classical one by 
a factor of $a^2(\eta)$, but the two loop correction to it would be down by a 
factor of $\alpha\ln(k/H)$. 

This seems a reasonable and probably provable conjecture. If it is true then
we can essentially get the full photon wave function by solving the 
integral-differential equation obtained from just the classical term and the 
one loop vacuum polarization in (\ref{SKeqn}). Which brings us to the second 
necessary extension of the current work: solving such an equation. Note that 
spatial translation invariance implies we can still use spatial plane waves. 
But then the spatial integrations are identical to the ones we already did in 
Section 5! We can therefore reduce the problem to one of solving for the 
multiplicative function of $\eta$. This is probably tractable analytically in 
some reasonable approximation. If not, it can certainly be done numerically.

Another issue is the extent to which our process repeats the Higgs mechanism of
flat space. We do not have Poincare invariance during inflation but one's
expectation is still that a massive photon has three polarizations. Since no
new degrees of freedom have been granted to the vector potential one might
suspect that the third polarization must come from the derivative of the 
scalar phase. It would be interesting to check this.

Finally, we comment on the possibility of important stochastic effects.
Although we have defined the photon wave function as the matrix element of
$A_{\mu}(x)$ between Bunch-Davies vacuum and a simultaneously prepared plane
wave photon state, one should bear in mind that the quantum averaging implicit
in matrix elements may give misleading results. The actual charge density 
induced by inflationary particle production is not smooth but rather stochastic
\cite{Linde}, corresponding to a highly squeezed state. About one massless, 
charged scalar exists per Hubble volume, moving at the speed of light in a 
random direction. We do not expect that this distinction amounts to a 
significant difference for super-horizon photons, which are affected by many 
different Hubble volumes. This can and should be checked.

\vskip 1cm
\centerline{\bf Acknowledgments}

We have profited from conversations with K. Dimopoulos and T. Vachaspati.
This work was partially supported by DOE contract DE-FG02-97ER\-41029 and 
by the Institute for Fundamental Theory at the University of Florida.

\end{document}